\documentclass[aps,prd,onecolumn,showpacs,superscriptaddress,preprintnumbers,floatfix,reprint,nofootinbib]{revtex4-1}
\usepackage{graphicx}
\usepackage{amsmath}
\usepackage{bm}
\usepackage{yhmath}
\usepackage[colorlinks,citecolor=blue,urlcolor=blue,linkcolor=blue]{hyperref}
\usepackage{subfigure}
\usepackage{color}
\usepackage{cases}
\usepackage{subfigure}
\usepackage{dcolumn,booktabs,bm}
\usepackage{slashed}
\usepackage{amsfonts,amssymb,stmaryrd,latexsym,amsmath}
\usepackage{textcomp}
\usepackage{multirow}
\usepackage{cancel}
\usepackage{diagbox}
\usepackage{array}

\renewcommand{\arraystretch}{1.8}
\allowdisplaybreaks

\begin{document}

\title{Higher fully-charmed tetraquarks:  Radial excitations and P-wave states}
\author{Guang-Juan Wang}
\affiliation{Advanced Science Research Center, Japan Atomic Energy
Agency, Tokai, Ibaraki, 319-1195, Japan}

\author{Lu Meng}\email{lu.meng@rub.de}
\affiliation{Institut f\"ur Theoretische Physik II, Ruhr-Universit\"at Bochum,  D-44780 Bochum,
 Germany}

\author{Makoto Oka}\email{oka@post.j-parc.jp}
\affiliation{Advanced Science Research Center, Japan Atomic Energy
Agency, Tokai, Ibaraki, 319-1195, Japan} \affiliation{Nishina Center
for Accelerator-Based Science, RIKEN, Wako 351-0198, Japan}

\author{Shi-Lin Zhu}\email{zhusl@pku.edu.cn}
\affiliation{School of Physics and Center of High Energy Physics,
Peking University, Beijing 100871,China}

\begin{abstract}
We systematically calculate the mass spectrum of the higher excited
fully-charmed tetraquark $cc\bar c\bar c$ states including the
S-wave radial excitations and the P-wave states with a
nonrelativistic quark model. The quark model is composed of a vector
one-gluon-exchange (OGE) and a scalar linear confinement interaction
with the parameters determined by the charmonium spectrum. In the
calculation, we consider both the $3_c-\bar 3_c$ and the
$6_c-\bar{6}_c$ color representations. For the $cc\bar c\bar c $
state, the $6_c-\bar 6_c$ component is located lower than the $\bar
3_c-3_c$ one because of the stronger attractive interactions between
the diquark and antidiquark. We focus on two excited modes and their
properties for the P-wave tetraquarks. The mass splitting for the
$\rho$-mode excitations with different color configurations is
large. The low-lying $6_c-\bar 6_c$ $\rho$-mode component helps to
explain the small mass gap between the ground S-wave and the P-wave
tetraquark states. The recently observed  $X(6900)$ state may be the
candidate of the first radially excited tetraquarks with
$J^{PC}=0^{++}$ or $2^{++}$, or the $1^{+-}$ or $2^{-+}$ P-wave
states based on the mass spectrum. Moreover, the lowest $T_c$ states with the exotic $J^{PC}$ quantum numbers $0^{--}$ and
$1^{-+}$ may decay into 
the P-wave $\eta_c J/\psi$ and di-$J/\psi$ modes,
respectively. The future
experimental search of these $T_c$ states will enrich the hadronic
spectrum.

\end{abstract}
\maketitle

\section{Introduction}\label{intro}
Since 2003, dozens of the exotic states named as XYZ states were
discovered in the mass range of the heavy quarkonium in experiments.
Amounts  of them cannot be categorized as  the conventional $Q\bar
Q$ ($Q=b,c$) mesons, and  are candidates  of the multi-quark states
$Q\bar Q q \bar q$ based on their quantum numbers (for example,
charge and $J^{PC}$) and decay channels. Various interpretations
have been proposed for their nature, including the hadronic
molecule, the compact tetraquark, the hado-charmonium, and so on
(for recent reviews, see
Refs.~\cite{Brambilla:2019esw,Chen:2016qju,Guo:2017jvc,Esposito:2016noz,Hosaka:2016pey,Ali:2017jda,Liu:2019zoy,Lebed:2016hpi}.).
However, none of the  exotic states have been firmly established so
far.

The light quark degrees of freedom in the XYZ state make the
investigation of their nature very complicated. Two heavy hadrons
can form the loosely bound molecules by exchanging light mesons such
as the deuteron ($P_c$ states are also the good candidates of
molecules~\cite{Meng:2019ilv,Wang:2019ato}). Alternatively, the four
quarks form a compact tetraquark state through the colored force.
Moreover, the  $Q\bar Q q \bar q$ states might strongly couple to
the $Q\bar Q$ core when their masses approach each
other~\cite{Eichten:2004uh}.

In contrast, the fully heavy tetraquark does not contain any light
quarks.  The light-meson-exchange interaction between two charmonia
is suppressed, while the heavy quarks could be bound through the
short-range colored forces arising from the gluon-exchange
interaction in QCD. Thus, they are more likely to be candidates of
the compact tetraquark states instead of the molecules. Moreover,
the fully heavy tetraquark states lie very far away from the
conventional heavy quarkonium states, which suppresses the
coupled-channel effect between the tetraquark components and the
$Q\bar Q$ core. Thus, the fully heavy tetraquark state becomes a
golden platform for the study of the multiquark states.

Before the experimental observation, there exist lots of theoretical
studies about the fully-heavy tetraquark states in
literature~\cite{Iwasaki:1975pv,Chao:1980dv,Ader:1981db,Zouzou:1986qh,Heller:1986bt,SilvestreBrac:1992mv}.
Recently, the CMS~\cite{Khachatryan:2016ydm,Durgut} and the
LHCb~\cite{Aaij:2018zrb} searched the fully-bottom state $X_{b b\bar
b\bar b}$ in the $\Upsilon(1S)\mu^+\mu^-$ channel. An analysis using
CMS data reported an excess in this channel~\cite{Durgut}. The
experimental processes inspired the intensive
discussions~\cite{Bai:2016int,Anwar:2017toa,Heupel:2012ua,Lloyd:2003yc,Debastiani:2017msn,
Barnea:2006sd,Berezhnoy:2011xy,Esposito:2018cwh,Karliner:2017qhf,Wu:2016vtq,Richard:2017vry,Liu:2019zuc,Czarnecki:2017vco,Hughes:2017xie,Chen:2016jxd,Berezhnoy:2011xn,Wang:2019rdo}.
These theoretical works focus on the existence of the stable bound
state below the lowest di-heavy-quarkonium threshold. Some
theoretical works support the existence of the bound
tetraquark~\cite{Anwar:2017toa,Heupel:2012ua,Bai:2016int,Lloyd:2003yc,Debastiani:2017msn,
Barnea:2006sd,Berezhnoy:2011xy,Esposito:2018cwh,Karliner:2017qhf},
while others do not favor it~\cite{Ader:1981db,Wu:2016vtq,
Liu:2019zuc,Hughes:2017xie,Richard:2017vry,Czarnecki:2017vco}.  Very
recently, the LHCb collaboration searched the di-$J/\psi$ channel
and reported a broad structure in the mass region $(6.2, 6.8)$ GeV
and a narrow  $X(6900)$ state with the signal significance more than
$5\sigma$~\cite{Aaij:2020fnh}. The experimental results inspired
numerous theoretical
discussions~\cite{Albuquerque:2020hio,liu:2020eha,Jin:2020jfc,Lu:2020cns,Giron:2020wpx,
    Dosch:2020hqm,Yang:2020wkh,Huang:2020dci,Hughes:2021xei,Faustov:2021hjs,Liang:2021fzr,Li:2021ygk,Guo:2020pvt,Dong:2020nwy,Wan:2020fsk,
    Zhu:2020snb,Feng:2020riv,Wang:2020gmd,Maciula:2020wri,Feng:2020qee,
    Goncalves:2021ytq,Huang:2021vtb,Mutuk:2021hmi,Lu:2020cns,Chen:2020xwe,
    Becchi:2020uvq,An:2020jix,Bedolla:2019zwg,Cao:2020gul,Chao:2020dml,Faustov:2020qfm,Gong:2020bmg,Gordillo:2020sgc,Karliner:2020dta,Ke:2021iyh, liu:2020eha,Lucha:2021mwx,Ma:2020kwb,Richard:2020hdw,Sonnenschein:2020nwn,Wang:2020dlo,Wang:2020wrp,Wang:2021xao,Weng:2020jao,Yang:2021hrb,Yang:2021zrc,Zhang:2020xtb,Zhao:2020cfi,Zhao:2020nwy,Zhu:2020xni}. The most popular interpretations are the compact tetraquark states~\cite{Albuquerque:2020hio,liu:2020eha,Jin:2020jfc,Lu:2020cns,Giron:2020wpx,Dosch:2020hqm,Yang:2020wkh,Huang:2020dci,Hughes:2021xei,Faustov:2021hjs,Liang:2021fzr,Li:2021ygk}. Other rare interpretations, such as coupled-channel effects of double-charmonium channels~\cite{Guo:2020pvt,Dong:2020nwy}, $c\bar c$ hybrid~\cite{Wan:2020fsk}, a Higgs-like boson~\cite{Zhu:2020snb}
and so on, are also proposed. The interpretations of their nature
are still in debate.  Other properties, for instance, the production
mechanisms~\cite{Feng:2020riv,Wang:2020gmd,Maciula:2020wri,Feng:2020qee,Goncalves:2021ytq,Huang:2021vtb,Mutuk:2021hmi}
and the decay patterns~\cite{Lu:2020cns,Chen:2020xwe,Becchi:2020uvq}
have also been discussed.

In our previous work before the observation of
$X(6900)$~\cite{Wang:2019rdo}, we calculated the fully heavy
tetraquark states.  We found that the ground S-wave $cc\bar c\bar c$
(noted as $T_c$ here and after)  states are located around $(6.3,
6.5)$ GeV. Meanwhile, we also predicted several radially excited
states around 6.9 GeV, which are consistent with the $X(6900)$
observed in LHCb. In Refs. \cite{Chen:2016jxd,liu:2020eha}, the
P-wave $T_c$ states were also predicted about $6.9$ GeV. For the
P-wave excited compact tetraquarks, one should include the
spin-orbital and tensor potentials in the quark model. Moreover, the
P-wave excitation may appear within the diquark/anti-diquark
clusters or between two clusters. Both the potentials and the
possible excitation modes  make the P-wave state much more
complicated than the S-wave ones.  In this work, we will refine and
extend our investigation on the fully-charmed $cc\bar c\bar c$
states in the compact tetraquark model. We will perform a dynamical
calculation for the mass spectrum of the higher tetraquark states,
including the S-wave radial excitations and the P-wave states with
two excitation modes.

To study the mass spectrum of the $T_c$ states, we adopt  the
nonrelativistic quark model with the minimal heavy quark interaction
which could describe charmonium spectrum. The potential contains the one-gluon-exchange (OGE) Coulomb-like, the linear confinement, and the hyperfine
interactions. The short-range spin-orbital and tensor potentials
arising from the OGE force and a long-range spin-orbital interaction
arising from the linear confinement term are treated perturbatively.
The color wave functions of the conventional mesons and baryons are
uniquely determined. However, there are two possible color
configurations for the  $T_c$ state, the $\bar 3_c-3_c$ and
$6_c-\bar 6_c$. In this work, we include both of the two color
configurations. With the careful treatment of the mixing effects, we
systematically calculate the $T_c$ mass spectra and investigate
their inner structures.

The paper is organized as follows. In Sec.~\ref{sec1}, we introduce
the Hamiltonian and construct the wave functions of the tetraquark
states. In Sec.~\ref{sec2} and Sec.~\ref{sec:pwv}, we present the
mass spectra and details for  S-wave and P-wave tetraquark states in
order. We provide the lower lying decay channels to search them in
Sec.~\ref{sec:decay}. In Sec.~\ref{sec3}, we adopt the harmonic
oscillator potential to qualitatively explain the small mass
splitting between the S-wave and P-wave $T_c$ states. Finally, we
give a brief  summary in Sec.~\ref{sec4}. In Appendix, we give the
spin, spin-orbital and tensor factors used in the calculation.

\section{Hamiltonian and wave function}\label{sec1}
For the fully heavy tetraquark state, we employ the non-relativistic
quark model proposed in Ref.~\cite{Barnes:2005pb}.The potentials are composed of the one-gluon-exchange potential $[G(q^2)(\gamma_\mu)_q\otimes (\gamma^\mu)_{\bar q}]$ and a phenomenological linear confinement potential $[S(Q^2)(1)_q (1)_{\bar q}]$ (where $G(Q^2)$ and $S(Q^2)$ are given in  Ref. \cite{Godfrey:1985xj}). The Hamiltonian
reads
\begin{eqnarray}
H & =&H_0+V^{(0)}+V^{(1)},\nonumber\\ H_0&=&\sum_{i=1}^{4}\frac{\mathbf{p}_{i}^{2}}{2m_{i}}+\sum_{i}m_{i}-T_G, \nonumber\\
V^{(0)}&=&\frac{{\mathbf \lambda}_i}{2}\cdot\frac{{\mathbf \lambda}_j}{2}V_{\text{cen}}(r_{ij}),\nonumber\\
V^{(1)}&=& \frac{{\mathbf \lambda}_i}{2}\cdot\frac{{\mathbf
\lambda}_j}{2}[V_\text{so}(r_{ij})+V_\text{tens}(r_{ij})],
\end{eqnarray}
where $\mathbf{p}_i$ and $m_i$ denote the momentum and mass of the
$i$th (anti)quark.  The kinematic energy of the system $T_G$
vanishes in the center mass system of the $T_c$ state. ${\mathbf
\lambda}_i$ is the color matrix. We divide the interaction into the
leading interaction $V^{(0)}$ and the perturbation $V^{(1)}$. The
leading  central potential $V_\text{cen}(r_{ij})$ is composed of the
OGE Coulomb-like $V_{\text{Coul}}$, linear confinement
$V_{\text{lin}}$ and screened hyperfine interactions
$V_{\text{hyp}}$,
\begin{eqnarray}
V_\text{cen}(r_{ij})&=&V_{\text{Coul}}+V_{\text{lin}}+V_{\text{hyp}}\nonumber \\
&=&\frac{\alpha_{s}}{r_{ij}}-\frac{3}{4}br_{ij}-\frac{8\pi\alpha_{s}}{3m_im_j}\left(\frac{\sigma}{\sqrt{\pi}}\right)^{3}e^{-\sigma^{2}r_{ij}^{2}}\mathbf{s}_{i}\cdot\mathbf{s}_{j}.\nonumber
\\ \label{eq:vcen}
\end{eqnarray}
For the orbitally excited states, the remaining perturbation,
spin-orbital potential $V_\text{so}$  and tensor potential
$V_{\text{tens}}$ may shift the mass spectrum obtained with the
leading interaction $V^{(0)}$. Their explicit forms are
\begin{eqnarray}
\label{eq:QM}
&&V_\text{so}(r_{ij})=V^v_\text{so}(r_{ij})+V^s_\text{so}(r_{ij}),\nonumber \\
&&V^v_\text{so}(r_{ij})=\frac{1}{r_{ij}}\frac{dV_{\text {Coul}}}{dr_{ij}}\frac{1}{4}\Big[\left(\frac{1}{m_{i}^{2}}+\frac{1}{m_{j}^{2}}+\frac{4}{m_{i}m_{j}}\right)\mathbf{L}_{ij}\cdot\mathbf{S}_{ij}\nonumber \\
&&~~~~~~~~~~+(\frac{1}{m_{i}^{2}}-\frac{1}{m_{j}^{2}})\mathbf{L}_{ij}\cdot(\mathbf{s}_{i}-\mathbf{s}_{j})\Big],\nonumber \\
&&V^s_\text{so}(r_{ij})=-\frac{1}{r_{ij}}\frac{dV_{\text{lin}}}{dr_{ij}}\left(\frac{\mathbf{L}_{ij}\cdot\mathbf{s}_{i}}{2m_{i}^{2}}+\frac{\mathbf{L}_{ij}\cdot\mathbf{s}_{j}}{2m_{j}^{2}}\right)\nonumber ,\\
&&V_{\text{tens}}(r_{ij})=-(\frac{\partial^{2}}{\partial
r_{ij}^{2}}-\frac{1}{r_{ij}}\frac{\partial}{\partial
r_{ij}})\frac{V_{\text{Coul}}}{3m_im_j}\mathcal{S}_{ij},\nonumber \\ \label{eq:vpert}
\end{eqnarray}
where $\mathbf {s}_i$ is the spin operator for the $i$th
(anti)quark. $\mathbf S_{ij}=\mathbf s_{i}+\mathbf s_{j}$ is the
spin operator for the $(ij)$th (anti)quark pair. The relative
orbital operators ${\mathbf L}_{ij}$ is defined as
\begin{equation}
    {\mathbf L}_{ij}=\mathbf{r}_{ij}\times \mathbf p_{ij}=\mathbf{r}_{ij} \times{m_i \mathbf p_i-m_j \mathbf p_i \over m_i+m_j}.
\end{equation}
The tensor operator $\mathcal{S}_{ij}$ reads
\begin{equation}
\mathcal{S}_{ij}=
\frac{3(\mathbf{s}_{i}\cdot\mathbf{r}_{ij})(\mathbf{s}_{j}\cdot\mathbf{r}_{ij})}{r_{ij}^{2}}-\mathbf{s}_{i}\cdot\mathbf{s}_{j}.
\end{equation}
The $V^v_\text{so}(r_{ij})$ and $V_{\text{tens}}(r_{ij})$ are
are given as relativistic corrections from the one-gluon-exchange potential as illustrated in Appendix A of Ref.~\cite{Godfrey:1985xj} and thus related to the
OGE Coulomb-like potential $V_{\text{Coul}}$. The
$V^s_\text{so}(r_{ij})$ arises from the linear confinement potential
$V_{\text{lin}}$.  In the calculation, we treat the potential 
$V_\text{cen}(r_{ij})$ as the leading effects and consider the
remaining  spin-orbital and tensor effects as the perturbation to
shift the mass spectrum.   We solve the Schr\"{o}dinger equation for the mesons and tetraquarks in two stages.  
 In the first stage, we  include the potential  $V^{(0)}$ in the Hamiltonian and solve the Schr\"{o}dinger  equation with the variational method to obtain the eigenstates.  In the second stage,  we consider the potential $V^{(0)}+V^{(1)}$
 in the Hamiltonian and  calculate the mass spectrum by diagonalizing the Hamiltonian matrix  in the basis  of eigenstates obtained in the first stage, where only the $V^{(0)}$ is considered.  As a benchmark, we calculate the mass
spectrum of the low-lying S-wave and P-wave charmonium with the
parameters in Ref.~\cite{Barnes:2005pb}. The results are summarized
in Table~\ref{tab:meson}. The quark model reproduces the
experimental values successfully. The deviations from the
experimental data are about $10$ MeV or even smaller, except that
for the third radically excited state $\psi(3S)$, which is $37$ MeV.

\begin{table}
 \renewcommand\arraystretch{1.5}
 \caption{The parameters of the quark  model and the corresponding mass spectrum (THE) of  the  charmonia $c\bar c$ compared with their experimental values (EXP)~\cite{Zyla:2020zbs}. }
 \label{tab:meson}
 \centering
 \setlength{\tabcolsep}{2.3mm}
\begin{tabular}{cccccc}
\toprule[1pt] \multicolumn{2}{c}{parameter} &
\multicolumn{4}{c}{Mass spectrum (MeV)}\tabularnewline \cline{3-6}
\multicolumn{2}{c}{} & $^{2S+1}L_{J}$ & Meson & EXP  &
THE\tabularnewline \midrule[1pt] $\alpha_{s}$ & 0.5461 & $^{1}S_{0}$
& $\eta_{c}$ & $2983.9$ & $2984$\tabularnewline

b [$\text{GeV}^2$] & 0.1452 & $^{3}S_{1}$ & $J/\psi$ & $3096.9$ &
$3092$\tabularnewline

$m_{c}$ {[}GeV{]} & 1.4794 & $^{3}P_{0}$ & $\chi_{c0}$ & $3414.7$ &
$3426$\tabularnewline

$\sigma$ {[}GeV{]} & 1.0946 & $^{3}P_{1}$ & $\chi_{c1}$ & $3510.7$ &
$3506$\tabularnewline

 &  & $^{1}P_{1}$ & $h_{c}(1P)$ & $3525.4$ & $3516$\tabularnewline

 &  & $^{3}P_{2}$ & $\chi_{c2}$ & $3556.2$ & $3556$\tabularnewline

 &  & $^{1}S_{0}$ & $\eta_{c}(2S)$ & $3637.5$ & $3634$\tabularnewline

 &  & $^{3}S_{1}$ & $\psi(2S)$ & $3686.1$ & $3675$\tabularnewline

 &  & $^{3}S_{1}$ & $\psi(3S)$ & $4039.0$  & $4076$\tabularnewline

 &  & $^{3}S_{1}$ & $\psi(4S)$ & $4421.0$ & $4412$\tabularnewline

\bottomrule[1pt]
\end{tabular}
\end{table}

\begin{figure}[htbp]
\centering
\includegraphics[width=0.5\textwidth]{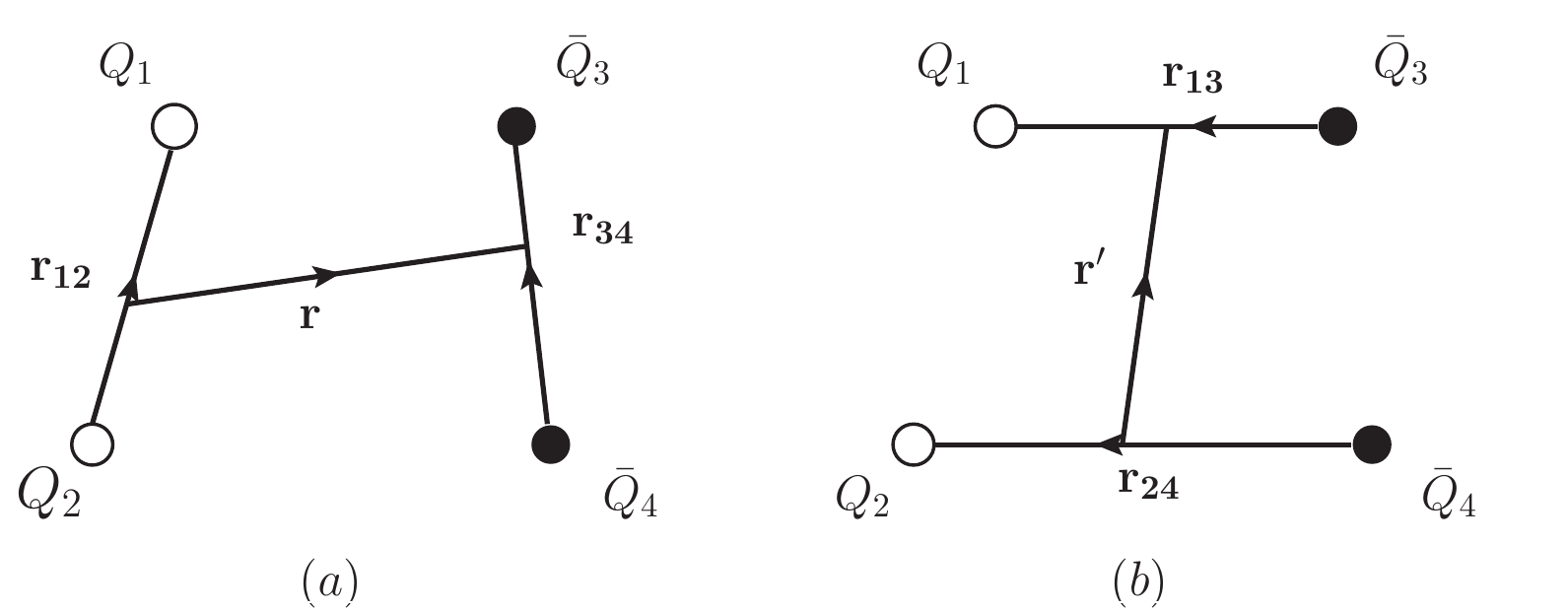}
\caption{Two sets of Jacobi coordinates in the four-quark system.
The $Q_1/Q_2$ and $\bar Q_3 /\bar Q_4$ are the quarks and
antiquarks, respectively.  An equivalent Jacobi coordinate to the
right one is the one under the permutation of  $Q_1\leftrightarrow
Q_2$ or  $\bar{Q}_3\leftrightarrow \bar{Q}_4$.} \label{fig:jac}
\end{figure}

The four-quark system $T_c$ can be described by three independent
coordinates shown in Fig.~\ref{fig:jac}. Corresponding to the two
sets of Jacobi coordinates, one has two methods to neutralize the
state in the color space, $|(Q_1Q_2)_{\bar 3_c}(\bar Q_3\bar
Q_4)_{3_c}\rangle$ and $|(Q_1Q_2)_{6_c}(\bar Q_3\bar Q_4)_{\bar
6_c}\rangle$ for the left diagram, while $|(Q_1\bar
Q_3)_{8_c}(Q_2\bar Q_4)_{8_c}\rangle$ and $|(Q_1\bar
Q_3)_{1_c}(Q_2\bar Q_4)_{1_c}\rangle$ ($Q_1\leftrightarrow Q_2$ or
$\bar Q_3\leftrightarrow \bar Q_4$) for the right one. The two sets
are equivalent to each other, which can be proved by the Fierz
transformation if one considers the complete spatial expansion
basis~\cite{Wang:2019rdo}. With the left set of Jacobi coordinates,
we can directly include the Fermi statistic constrains on the quarks
and antiquarks as listed in Table~\ref{tab:cfs}, which simplifies
the calculation.  Thus, we expand the tetraquark state with the
Gaussian wave functions in the basis of the left Jacobi
coordinates~\cite{Hiyama:2003cu,Hiyama:2018ivm}. The wave functions
of the  $T_c$ state with the angular momentum $(J,J_z)$ read,
\begin{eqnarray}\label{wavefunction}
&&\Psi_{JJ_z}=\sum_{\alpha}A_{\alpha}
\cdot \chi_c\phi_{n_{a}l_{a}}(r_{12},\beta_{a})\phi_{n_{b}l_{b}}(r_{34},\beta_{b})\phi_{n_{ab}l_{ab}}(r,\beta)\nonumber\\
&&  \cdot \left[[\left(Y_{l_{a}}(\hat{r}_{12})\otimes\chi_{s_{a}}\right)_{j_{a}}\left(Y_{l_{b}}(\hat{r}_{34})\otimes\chi_{s_{b}}\right)_{j_{b}}]_j\otimes Y_{l_{ab}}(\hat{r})\right]_{JJ_{z}},\nonumber\\
\end{eqnarray}
where the spatial wave function reads
\begin{eqnarray}
\phi_{n_a l_a}({r}_{12},\beta_a)&=&\Big\{
\frac{2^{l_a+2}(2\nu_{n_a})^{l+3/2}}{\sqrt \pi (2l_a+1)!!}\Big
\}^{1/2}r_{12}^{l_{a}} e^{-\nu_{n_a} r_{12}^{2}},\nonumber
\end{eqnarray}
with $\nu_{n_a}$ being the oscillating parameter,
\begin{eqnarray}
\nu_{n_a}=\frac{n_a\beta^2_{a}}{2}, \,\,\,
(n_a=1,2,..,n^{\text{max}}_a),
\end{eqnarray}
The $n_a/n_b/n_{ab}$ are related to the radial excitations along the
three relative coordinates. $l_a$/$l_b$/$l_{ab}$ are the orbital
angular momenta inside the diquark/antidiquarks, and that between
the two clusters, respectively. The spin of the diquark
(antidiquarks) $s_a$ ($s_b$) couples with  $l_a$ ($l_b$) into the $
j_a$ ($ j_b$).~$j_a$ and $j_b$ couple into $j$ and then combine with
$l_{ab}$ to form the total angular momentum $J$. The $\alpha$
represents the basis with the quantum number $\{ n_a, l_a, s_a, j_a,
n_b, l_b, s_b, j_b, n_{ab},l_{ab},j\}$ which can form the angular
momentum $J$. The expanding parameter  $A_{\alpha}$ and the
oscillating parameters  $\beta_{a/b/ab}$ are determined by solving
the Schr\"{o}dinger equation with the variational method.

The $\chi_s$ and $\chi_c$ stand for  the spin and color wave
functions, respectively. We present the possible color-flavor-spin
configurations of the $T_c$ state in  Table~\ref{tab:csfwf}.  For
the P-wave state, there are two orbital excitation modes, the
$\rho$-mode state with the orbital excitation in the diquark or
antidiqaurk, i.e., $|l_a=1, l_b=l_{ab}=0\rangle$ or
$|l_a=0,l_b=1,l_{ab}=0\rangle$, and the $\lambda$-mode one with the
orbital excitation between the two clusters, i.e.,
$|l_a=l_b=0,l_{ab}=1\rangle$. In general, the state should be a
superposition of the  two modes.

\begin{table*}
 \renewcommand\arraystretch{1.5}
  \caption{The color-flavor-spin configurations of the $QQ$ ($\bar Q \bar Q$) diquark (antidiquark). The scripts ``S" and ``A"  represent the exchange symmetry and antisymmetry for the identical particles, respectively.}\label{tab:cfs}
 \centering
 \setlength{\tabcolsep}{2.3mm}
\begin{tabular}{c|c|cccc}
\toprule[1pt] Flavor & S-wave($L=0$) & Spin & Color &  &
$J^{P}$\tabularnewline \midrule[1pt] S & S & S($S_{QQ}=1$) &
$\bar{3}_{c}$(A) & $[QQ]_{\bar{3}_{c}}^{1}$ & $1^{+}$\tabularnewline
S & S & A($S_{QQ}=0$) & $6_{c}$(S) & $[QQ]_{6_{c}}^{0}$ &
$0^{+}$\tabularnewline \midrule[1pt] Flavor & P-wave ($L=1$) & Spin
& Color &  & \tabularnewline \midrule[1pt] \multirow{3}{*}{S} &
\multirow{3}{*}{A} & \multirow{3}{*}{S($S_{QQ}=1$)} &
\multirow{3}{*}{$6_{c}$(S)} &
$\left[[QQ]_{6_{c}}^{1},\rho\right]_{6_{c}}^{0}$ &
$0^{-}$\tabularnewline

 &  &  &  & $\left[[QQ]_{6_{c}}^{1},\rho\right]_{6_{c}}^{1}$ & $1^{-}$\tabularnewline

 &  &  &  & $\left[[QQ]_{6_{c}}^{1},\rho\right]_{6_{c}}^{2}$ & $2^{-}$\tabularnewline
\hline S & A & A($S_{QQ}=0$) & $\bar{3}_{c}$(A) &
$\left[[QQ]_{\bar{3}_{c}}^{0},\rho\right]_{\bar{3}_{c}}^{1}$ &
$1^{-}$\tabularnewline

\bottomrule[1pt]
\end{tabular}
\end{table*}

\begin{table*}
 \renewcommand\arraystretch{2.5}
 \caption{The color-flavor-spin wave functions of  the S-wave and P-wave tetraquark states with different $J^{PC}$. The superscripts and subscripts in the wave functions represent the spin and the color representations, respectively. We use $|c^C_{\rho/\lambda};^{2S+1}P_J\rangle$ to label P-wave states, where $c=$ 3 or 6 stand for the color representation and $C$ is the C-parity of the system.  }
 \label{tab:csfwf}
 \centering
 \setlength{\tabcolsep}{2.5mm}
\begin{tabular}{c|ccccc}
\toprule[1pt] \multirow{2}{*}{S-wave}  & \multirow{2}{*}{$0^{++}$} &
\multicolumn{1}{c}{$\left[[QQ]_{\bar{3}_{c}}^{1}[\bar{Q}\bar{Q}]_{3_{c}}^{1}\right]_{1_{c}}^{0}$}
& $1^{+-}$ &
$\left[[QQ]_{\bar{3}_{c}}^{1}[\bar{Q}\bar{Q}]_{3_{c}}^{1}\right]_{1_{c}}^{1}$\tabularnewline
 &  & \multicolumn{1}{c}{$\left[[QQ]_{6_{c}}^{0}[\bar{Q}\bar{Q}]_{\bar{6}_{c}}^{0}\right]_{1_{c}}^{0}$} & $2^{++}$ & $\left[[QQ]_{\bar{3}_{c}}^{1}[\bar{Q}\bar{Q}]_{3_{c}}^{1}\right]_{1_{c}}^{2}$\tabularnewline
\hline
 &  & \multicolumn{1}{c}{$\lambda$-mode} &  & $\rho$-mode\tabularnewline
 \cline{2-5}
\multirow{14}{*}{P-wave} & \multirow{2}{*}{$0^{-+}$} &
\multirow{2}{*}{$|3^{+}_\lambda;
^{3}P_{0}\rangle=\left[\left[[QQ]_{\bar{3}_{c}}^{1}[\bar{Q}\bar{Q}]_{3_{c}}^{1}\right]_{1_{c}}^{1},\lambda\right]_{1_{c}}^{0}$}
& \multirow{14}{*}{} & $|3^{+}_{\rho};
^3P_0\rangle=\frac{1}{\sqrt{2}}\left(\left[\left[[QQ]_{\bar{3}_{c}}^{0},\text{\ensuremath{\rho}}\right]_{\bar{3}_{c}}^{1}[\bar{Q}\bar{Q}]_{3_{c}}^{1}\right]^{0}+c.c.\right)$\tabularnewline

 &  &  &  & $|6^{+}_\rho; ^{3}P_{0}\rangle=\frac{1}{\sqrt{2}}\left(\left[\left[[QQ]_{6_{c}}^{1},\rho\right]_{6_{c}}^{0}[\bar{Q}\bar{Q}]_{\bar{6}_{c}}^{0}\right]^{0}+c.c.\right)$\tabularnewline
 \cline{2-5}
  & \multirow{2}{*}{$1^{-+}$} & \multirow{2}{*}{$\ensuremath{|3^{+}_\lambda; ^{3}P_{1}\rangle=\left[\left[[QQ]_{\bar{3}_{c}}^{1}[\bar{Q}\bar{Q}]_{3_{c}}^{1}\right]_{1_{c}}^{1},\lambda\right]_{1_{c}}^{1}}$} &  & $|3^{+}_\rho; ^{3}P_{1}\rangle=\frac{1}{\sqrt{2}}\left(\left[\left[[QQ]_{\bar{3}_{c}}^{0},\text{\ensuremath{\rho}}\right]_{\bar{3}_{c}}^{1}[\bar{Q}\bar{Q}]_{3_{c}}^{1}\right]^{1}+c.c.\right)$\tabularnewline

 &  &  &  & $|6^{+}_\rho; ^{3}P_{1}\rangle=\frac{1}{\sqrt{2}}\left(\left[\left[[QQ]_{6_{c}}^{1},\rho\right]_{6_{c}}^{1}[\bar{Q}\bar{Q}]_{\bar{6}_{c}}^{0}\right]^{1}+c.c.\right)$\tabularnewline
 \cline{2-5}
 & \multirow{2}{*}{$2^{-+}$} & \multirow{2}{*}{$\ensuremath{|3^{+}_\lambda; ^{3}P_{2}\rangle=\left[\left[[QQ]_{\bar{3}_{c}}^{1}[\bar{Q}\bar{Q}]_{3_{c}}^{1}\right]_{1_{c}}^{1},\lambda\right]_{1_{c}}^{2}}$} &  & $|3^{+}_\rho; ^{3}P_{2}\rangle=\frac{1}{\sqrt{2}}\left(\left[\left[[QQ]_{\bar{3}_{c}}^{0},\text{\ensuremath{\rho}}\right]_{\bar{3}_{c}}^{1}[\bar{Q}\bar{Q}]_{3_{c}}^{1}\right]^{2}+c.c.\right)$\tabularnewline

 &  &  &  & $|6^{+}_\rho; ^{3}P_{2}\rangle=\frac{1}{\sqrt{2}}\left(\left[\left[[QQ]_{6_{c}}^{1},\rho\right]_{6_{c}}^{2}[\bar{Q}\bar{Q}]_{\bar{6}_{c}}^{0}\right]^{2}+c.c.\right)$\tabularnewline
\cline{2-5}
 & \multirow{2}{*}{$0^{--}$} & \multirow{2}{*}{$-$} &  & $|3^{-}_\rho; ^{3}P_{0}\rangle=\frac{1}{\sqrt{2}}\left(\left[\left[[QQ]_{\bar{3}_{c}}^{0},\text{\ensuremath{\rho}}\right]_{\bar{3}_{c}}^{1}[\bar{Q}\bar{Q}]_{3_{c}}^{1}\right]^{0}-c.c.\right)$\tabularnewline

 &  &  &  & $|6^{-}_\rho; ^{3}P_{0}\rangle=\frac{1}{\sqrt{2}}\left(\left[\left[[QQ]_{6_{c}}^{1},\rho\right]_{6_{c}}^{0}[\bar{Q}\bar{Q}]_{\bar{6}_{c}}^{0}\right]^{0}-c.c.\right)$\tabularnewline
\cline{2-5}
 & \multirow{3}{*}{$1^{--}$} & $|3^{-}_\lambda; ^{1}P_{1}\rangle=\left[\left[[QQ]_{\bar{3}_{c}}^{1}[\bar{Q}\bar{Q}]_{3_{c}}^{1}\right]_{1_{c}}^{0},\lambda\right]_{1_{c}}^{1}$ &  & $|3^{-}_\rho; ^{3}P_{1}\rangle=\frac{1}{\sqrt{2}}\left(\left[\left[[QQ]_{\bar{3}_{c}}^{0},\text{\ensuremath{\rho}}\right]_{\bar{3}_{c}}^{1}[\bar{Q}\bar{Q}]_{3_{c}}^{1}\right]^{1}-c.c.\right)$\tabularnewline

 &  & $|6^{-}_\lambda; ^{1}P_{1}\rangle=\left[\left[[QQ]_{6_{c}}^{0}[\bar{Q}\bar{Q}]_{\bar{6}_{c}}^{0}\right]_{1_{c}}^{0},\lambda\right]_{1_{c}}^{1}$ &  & $|6^{-}_\rho; ^{3}P_{1}\rangle=\frac{1}{\sqrt{2}}\left(\left[\left[[QQ]_{6_{c}}^{1},\rho\right]_{6_{c}}^{1}[\bar{Q}\bar{Q}]_{\bar{6}_{c}}^{0}\right]^{1}-c.c.\right)$\tabularnewline

 &  & $|3^{-}_\lambda; ^{5}P_{1}\rangle=\left[\left[[QQ]_{\bar{3}_{c}}^{1}[\bar{Q}\bar{Q}]_{3_{c}}^{1}\right]_{1_{c}}^{2},\lambda\right]_{1_{c}}^{1}$ &  & \tabularnewline
\cline{2-5}
 &  \multirow{2}{*}{$2^{--}$}  & \multirow{2}{*}{$|3^{-}_\lambda; ^{5}P_{2}\rangle=\left[\left[[QQ]_{\bar{3}_{c}}^{1}[\bar{Q}\bar{Q}]_{3_{c}}^{1}\right]_{1_{c}}^{2},\lambda\right]_{1_{c}}^{2}$} &  & $|3^{-}_\rho; ^{3}P_{2}\rangle=\frac{1}{\sqrt{2}}\left(\left[\left[[QQ]_{\bar{3}_{c}}^{0},\text{\ensuremath{\rho}}\right]_{\bar{3}_{c}}^{1}[\bar{Q}\bar{Q}]_{3_{c}}^{1}\right]^{2}-c.c.\right)$\tabularnewline

 &  &  &  & $|6^{-}_\rho; ^{3}P_{2}\rangle=\frac{1}{\sqrt{2}}\left(\left[\left[[QQ]_{6_{c}}^{1},\rho\right]_{6_{c}}^{2}[\bar{Q}\bar{Q}]_{\bar{6}_{c}}^{0}\right]^{2}-c.c.\right)$\tabularnewline
\cline{2-5}
 & $3^{--}$ & $|3^{-}_\lambda; ^{5}P_{3}\rangle=\left[\left[[QQ]_{\bar{3}_{c}}^{1}[\bar{Q}\bar{Q}]_{3_{c}}^{1}\right]_{1_{c}}^{2},\lambda\right]_{1_{c}}^{3}$ &  & \tabularnewline
 \bottomrule[1pt]
\end{tabular}
\end{table*}

\section{Spectrum of S-wave tetraquarks and their radial excitations} \label{sec2}
The possible spin and parity  quantum numbers of the S-wave $T_c$
states are $J^{PC}=0^{++}$, $1^{+-}$, and $2^{++}$.~An S-wave
tetraquark state may couple with the orbitally  excited tetraquarks
with at least two orbital angular momentum, i.e.,
$(l_a,l_b,l_{ab})=(1,1,0), (0,1,1), (1,1,2)$, etc. For the $T_c$
state, such orbital excited states will couple with the S-wave ones
through spin-orbital and tensor potentials in Eq.~(\ref{eq:vpert}),
which could raise or lower the orbital momentum. Such interactions
contribute to the mass shifting  slightly. Thus, we do not include
such high orbital excitations for the S-wave tetraquarks and
concentrate on the state without any orbital angular momentum.

\begin{figure}[htbp]
\centering
\includegraphics[width=0.5\textwidth]{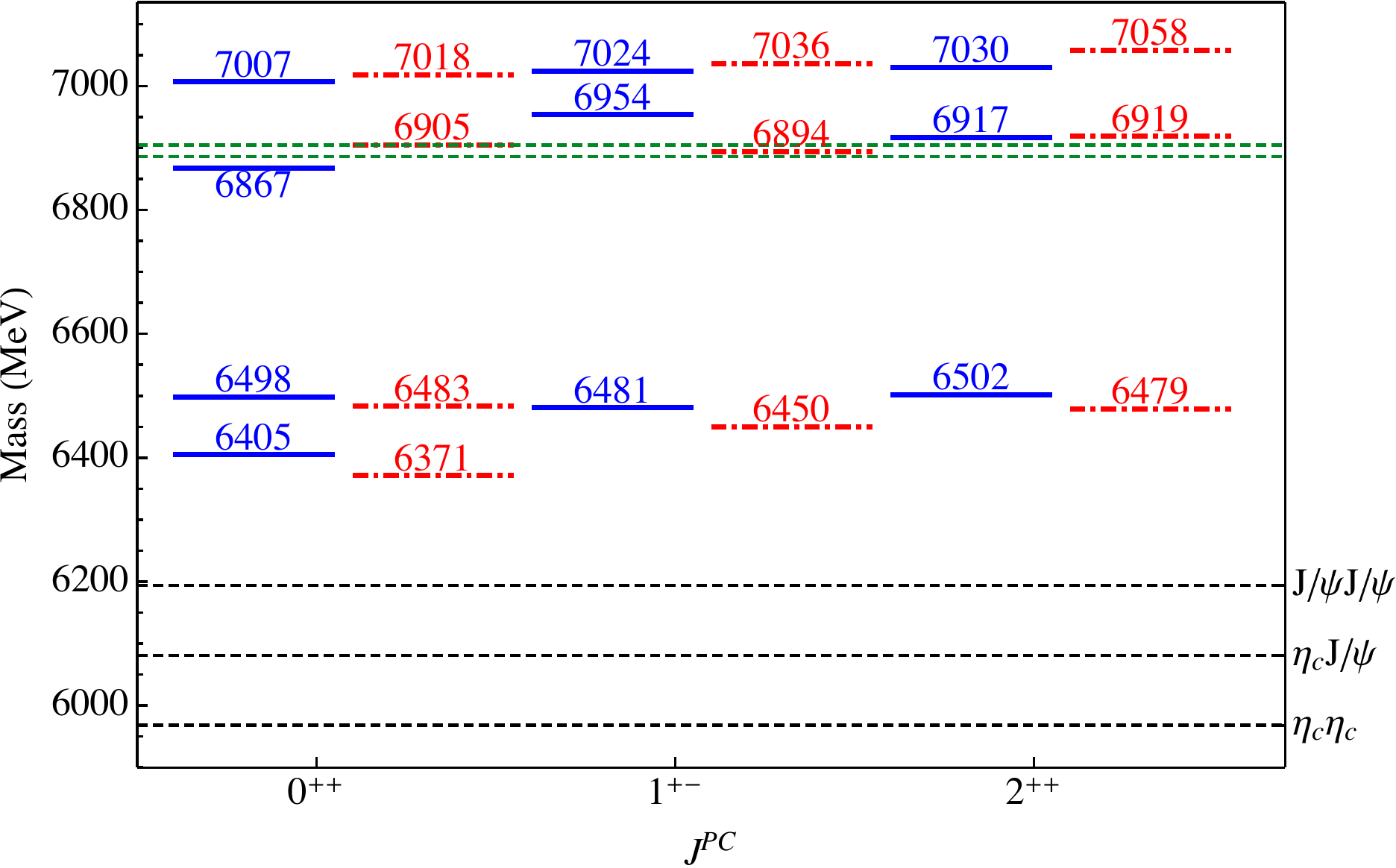}
\caption{The mass spectrum of the S-wave tetraquark states $T_c$.
The dot-dashed red and the blue bars represent the mass spectra from
the quark model in Ref.~\cite{Silvestre-Brac:1996myf} and
Ref.~\cite{Barnes:2005pb}, respectively. The green dashed lines
stand for $6886$ MeV and $6905$ MeV, which are the central values of
the $X(6900)$ mass in the two fits obtained by
experiments~\cite{Aaij:2020fnh}.} \label{fig:Swave}
\end{figure}

We present the results in Fig.~\ref{fig:Swave} with the blue bars.
For comparison, we also draw the results with the dot-dashed red
bars using another quark model proposed in
Ref.~\cite{Silvestre-Brac:1996myf}, which has been employed in our
previous work~\cite{Wang:2019rdo}. In Fig.~\ref{fig:Swave}, one
finds that the two sets of numerical results are similar to each
other within tens of MeV for the low-lying excited states. For the
higher radially exited states, the differences will become larger.
This arises from the unavoidable uncertainties of quark model in the
sector. Furthermore,  the higher S-wave radially excited states might be located very close to the non-S-wave excited states, such as $(l_a,l_b,l_{ab})=(1,1,2)$. The spin-orbital and tensor potentials will lead to their mixing effects. If their mass difference is small, even small contributions  from the spin-orbital and tensor potentials may lead to large mixing. This will influence both the mass spectrum and the decay patterns. The mixing effect therefore cannot be neglected.  Our present calculations are still robust for
the low-lying tetraquark states. Additionally, the lowest $T_c$
states are located above the lowest two-charmonium channels. We
therefore conclude that there is no bound state for the $T_c$
states, which is consistent with our conclusions in the previous
work~\cite{Wang:2019rdo}. We also display the details of the S-wave
states in Table~\ref{tab:swave}, including the radius and the
components of different color configurations. According to the
radius, the four (anti)-quarks are confined into a compact state.

\begin{table*}
 \renewcommand\arraystretch{1.5}
  \caption{The mass spectrum (MeV), the percentage of different color configurations, and the root mean square radius (fm) of the S-wave tetraquark states. }\label{tab:swave}
 \centering
 \setlength{\tabcolsep}{2.5mm}
\begin{tabular}{c|c|cccccccc}
\toprule[1pt]
$0^{++}$ & Mass & $\bar{3}_{c}\otimes3_{c}$ &
$6_{c}\otimes\bar{6}_{c}$ & $1_{c}\otimes1_{c}$ &
$8_{c}\otimes8_{c}$ & $r_{12}/r_{34}$ & $r$ & $r_{13}/r_{24}$ &
$r'$\tabularnewline \hline

\multirow{2}{*}{1S} & $6405$ & $31.9\%$ & $68.1\%$ & $96.9\%$ &
$3.13\%$ & $0.52$ & $0.31$ &  $0.48$ & $0.37$ \tabularnewline
  & $6498$ & $67.7\%$ & $32.3\%$ & $5.7\%$ & $94.3\%$ & $0.51$ & $0.36$ & $0.51$& $0.36$ \tabularnewline

\multirow{2}{*}{2S} & $6867$ & $10.6\%$ & $89.4\%$ & $80.6\%$ &
$19.4\%$ & $0.65$ & $0.35$ &  $0.58$ & $0.46$ \tabularnewline

 & $7007$ & $89.7\%$ & $10.3\%$ & $26.0\%$ & $74.0\%$ & $0.49$ & $0.47$ &  $0.59$ & $0.35$ \tabularnewline

\hline $1^{+-}$ & Mass & $\bar{3}_{c}\otimes3_{c}$ &
$6_{c}\otimes\bar{6}_{c}$ & $1_{c}\otimes1_{c}$ &
$8_{c}\otimes8_{c}$ & $r_{12}/r_{34}$ & $r$ & $r_{13}/r_{24}$ &
$r'$\tabularnewline \hline 1S & $6481$ & 100\% & 0\% & $33.3\%$ &
$66.7\%$ & $0.48$ & $0.37$  & $0.51$ & $0.34$  \tabularnewline

2S & $6954$ & 100\% & 0\% & $33.3\%$ & $66.7\%$ & $0.61$ & $0.44$  & $0.61$ &
$0.43$  \tabularnewline

3S & $7024$ & 100\% & 0\% & $33.3\%$ & $66.7\%$ & $0.66$ & $0.42$  & $0.62$  &
$0.46$ \tabularnewline \hline

$2^{++}$& Mass & $\bar{3}_{c}\otimes3_{c}$ &
$6_{c}\otimes\bar{6}_{c}$ & $1_{c}\otimes1_{c}$ &
$8_{c}\otimes8_{c}$ & $r_{12}/r_{34}$ & $r$ & $r_{13}/r_{24}$ &
$r'$\tabularnewline

\hline 1S & $6502$ & 100\% & 0\% & $33.3\%$ & $66.7\%$ & $0.49$ &
$0.39$ & $0.53$ & $0.35$ \tabularnewline
 2S & $6917$ & 100\% & 0\%& $33.3\%$ & $66.7\%$ & $0.55$ & $0.60$ & $0.72$ & $0.39$ \tabularnewline
 3S & $7030$ & 100\% & 0\%& $33.3\%$ & $66.7\%$ & $0.64$ & $0.46$& $0.64$  & $0.45$ \tabularnewline
\bottomrule[1pt]
\end{tabular}
\end{table*}

\section{P-wave tetraquark spectrum} \label{sec:pwv}

For the P-wave tetraquark, we consider the $\rho$-mode and the
$\lambda$-mode tetraquarks with one orbital excitation. First, we
calculate the eigenstates using the leading interactions, including
the OGE Coulomb-like plus  linear confinement, and the hyperfine
potentials in Eq.~\eqref{eq:vcen}.

As illustrated in Table~\ref{tab:csfwf}, the color-flavor-spin wave
functions  of the $T_c$ states with  $J^{PC}=J^{-+}$ ($J=0,1, 2$)
are the same except  for the total couplings of the spin and orbital
angular momentum. With the leading Coulomb, confinement and
hyperfine potentials, their mass spectrum are also the same. The
mass difference comes from the perturbative spin-orbital or tensor
potentials.  This also applies to the $T_c$'s with the negative
charge conjugate parity $J^{--}$ ($J=1,2,3$) except for the $1^{--}$
one, which has two extra $\lambda$-mode states. We list the
eigenstates and corresponding percentages of different components
under the leading potentials in Table~\ref{tab:lambdarholo}.

In the calculation, we find that the color electric interactions do
not induce the mixing of the $\lambda$- and $\rho$-mode $T_c$
excitations. From Table~\ref{tab:csfwf}, we can see that the spin
wave functions are orthogonal for the $\lambda$- and $\rho$-mode
constrained by the Fermi-Dirac statistics. Thus, the electric
interactions do not mix the $\lambda$- and $\rho$-mode.

Next, we concentrate on the hyperfine interaction with the color
magnetic operator. For the $J^{PC}=1^{--}$ and $J^{PC}=2^{--}$
tetraquarks, the total spin of the $\lambda$-mode states is $0$ or
$2$ as shown in Table~\ref{tab:csfwf}, while that of  the
$\rho$-mode state is $1$. The hyperfine potential cannot flip the
total spin as
\begin{eqnarray}
\langle
[{\chi^{s_a}_s}\otimes{\chi^{s_b}_s}]^{S_{\lambda}=0/2}|\mathbf{s}_{i}\cdot\mathbf{s}_{j}|[{\chi^{s'_a}_s}\otimes
{\chi^{s'_b}_s}]^{S_{\rho}=1}\rangle=0,
\end{eqnarray}
where the subscripts $a$ and $b$ represent the diquark and
antidiquark, respectively. For the $J^{PC}=0^{-+}$, $1^{-+}$,
$2^{-+}$ tetraquarks, the hyperfine potentials will contribute to
the mixing of the two P-wave excitation modes. The mixing is quite
small and the eigenstates are nearly total $\lambda$ or $\rho$-mode
excitations. We denote them with their main excitation modes.

In Table~\ref{tab:lambdarholo}, we obtain the relationship between
the $\rho$- and $\lambda$-mode excitations with different color
configurations as $6_{\rho}<3_{\lambda}<6_{\lambda}<3_{\rho}$ (the
number and the subscript represent the color configuration and the
excited mode, respectively). For the $\rho$-mode $T_c$ eigenstates,
the color configurations of the lower  $|\rho_{1}^{+/-}\rangle$
state and the higher $|\rho^{+/-}_{2}\rangle$ are dominated by
$6_c-\bar 6_c$ and $\bar 3_c-3_c$ components, respectively. Similar
to the S-wave tetraquarks, the $6_c-\bar 6_c$  component lies lower
than the $\bar 3_c-3_c$ one. In the $6_c-\bar 6_c$ configuration,
although the interactions in the diquark or antidiquark are
repulsive, the potentials between the two clusters are attractive
and much stronger than those in the $\bar 3_c-3_c$ one. This  leads
to a confined $6_c-\bar 6_c$ state, which is even lower than the
$\bar 3_c-3_c$ one in our quark model.

What is more, as shown in Table \ref{tab:lambdarholo}, the mass
difference between the lower $|\rho_{1}^{+/-}\rangle$ and  higher
$|\rho^{+/-}_{2}\rangle$ is about $300$ MeV, while that between the
two ground $0^{++}$ S-wave $T_c$ states is about $100$ MeV. This
indicates that the mass splittings between the $6_c-\bar 6_c$ and
$3_c-\bar 3_c$ $\rho$-mode states should be much larger, which is
useful to investigate the complicated  color configurations in the
multi-quark state, especially  the rarely known $6_c-\bar 6_c$ color
configurations.

\begin{table*}
 \renewcommand\arraystretch{1.5}
  \caption{The mass spectrum (MeV) and the percentages of different color configurations in the P-wave  $T_c$ states obtained with leading potentials,  including the Coulomb, linear confinement and  hyperfine potentials in Eq.~\eqref{eq:vcen}. The eigenstates are labeled by $|\rho/\lambda ^{C}_i\rangle$,where $C$ is the C-parity and $i=1,2,3$ represent different states in the ascending order of the mass.  }\label{tab:lambdarholo}
 \centering
 \setlength{\tabcolsep}{2.5mm}
\begin{tabular}{c|c|ccc|cc|c}
  \toprule[1pt]
    $J^{-+}$ & Mass & $|{3}_\lambda^{+};^{3}P_{0,1,2}\rangle$& $|3_{\rho}^{+};{}^{3}P_{0,1,2}\rangle$ & $|6_{\rho}^{+};{}^{3}P_{0,1,2}\rangle$  & $1_{c}\otimes1_{c}$ & $8_{c}\otimes8_{c}$ & \multirow{5}{*}{$6_{\rho}^{+}<3_{\lambda}^{+}<3_{\rho}^{+}$}\tabularnewline
    $|\lambda_1 ^+\rangle$ & 6746 & $99.5\%$ & $0.4\%$  &  $0.1\%$ &$33.4\%$  &$66.6\%$  & \tabularnewline
    \cline{1-7} \cline{2-7} \cline{3-7} \cline{4-7} \cline{5-7} \cline{6-7} \cline{7-7}
    $J^{-+}$ & Mass & $|{3}_\lambda^{+};^{3}P_{0,1,2}\rangle$ &$|3_{\rho}^{+};{}^{3}P_{0,1,2}\rangle$ & $|6_{\rho}^{+};{}^{3}P_{0,1,2}\rangle$  & $1_{c}\otimes1_{c}$ & $8_{c}\otimes8_{c}$ & \tabularnewline
    $|\rho_{1}^+\rangle$ & 6599 &$0.1\%$ & $24.5\%$ & $75.4\%$   & $58.5\%$ & $41.5\%$ \tabularnewline
    $|\rho_{2}^+\rangle$ & 6894  & $0.5\%$& $72.0\%$ & $27.5\%$ &   $42.5\%$ & $57.5\%$ & \tabularnewline
    \hline
    $J^{--}$ & Mass & $|3_{\lambda}^{-};^{1}P_{1}\rangle$ & $|6_{\lambda}^{-};^{1}P_{1}\rangle$ & $|3_{\lambda}^{-};{}^{5}P_{1,2,3}\rangle$ & $1_{c}\otimes1_{c}$ & $8_{c}\otimes8_{c}$ & \multirow{7}{*}{$6_{\rho}^{-}<3_{\lambda}^{-}<6_{\lambda}^{-}<3_{\rho}^{-}$}\tabularnewline
    $|\lambda_{1}^-\rangle$ & 6740 & $98.9\%$ & $1.1\%$ & $0\%$ & $33.7\%$ & $66.3\%$  & \tabularnewline
    $|\lambda_{2}^-\rangle$ & 6741 & $0\%$ & $0\%$ & $100\%$ &$33.3\%$  &$66.7\%$  & \tabularnewline
    $|\lambda_{3}^-\rangle$ & 6885 & $1.4\%$ & $98.6\%$ & $0\%$ & $66.2\%$  &$33.8\%$  & \tabularnewline
    \cline{1-7} \cline{2-7} \cline{3-7} \cline{4-7} \cline{5-7} \cline{6-7} \cline{7-7}
    $J^{--}$ & Mass & $|3_{\rho}^{-};^{3}P_{0,1,2}\rangle$ & $|6_{\rho}^{-};^{3}P_{0,1,2}\rangle$ &  & $1_{c}\otimes1_{c}$ & $8_{c}\otimes8_{c}$ & \tabularnewline
    $|\rho_{1}^-\rangle$ & 6561 & $27.1\%$ & $72.9\%$ &  & $57.6\%$ & $42.4\%$ & \tabularnewline
    $|\rho_{2}^-\rangle$ & 6913 & $72.1\%$ & $27.9\%$ &  & $42.6\%$ & $57.4\%$ & \tabularnewline
  \bottomrule[1pt]
\end{tabular}
\end{table*}

\begin{figure}[htbp]
\centering
\includegraphics[width=0.5\textwidth]{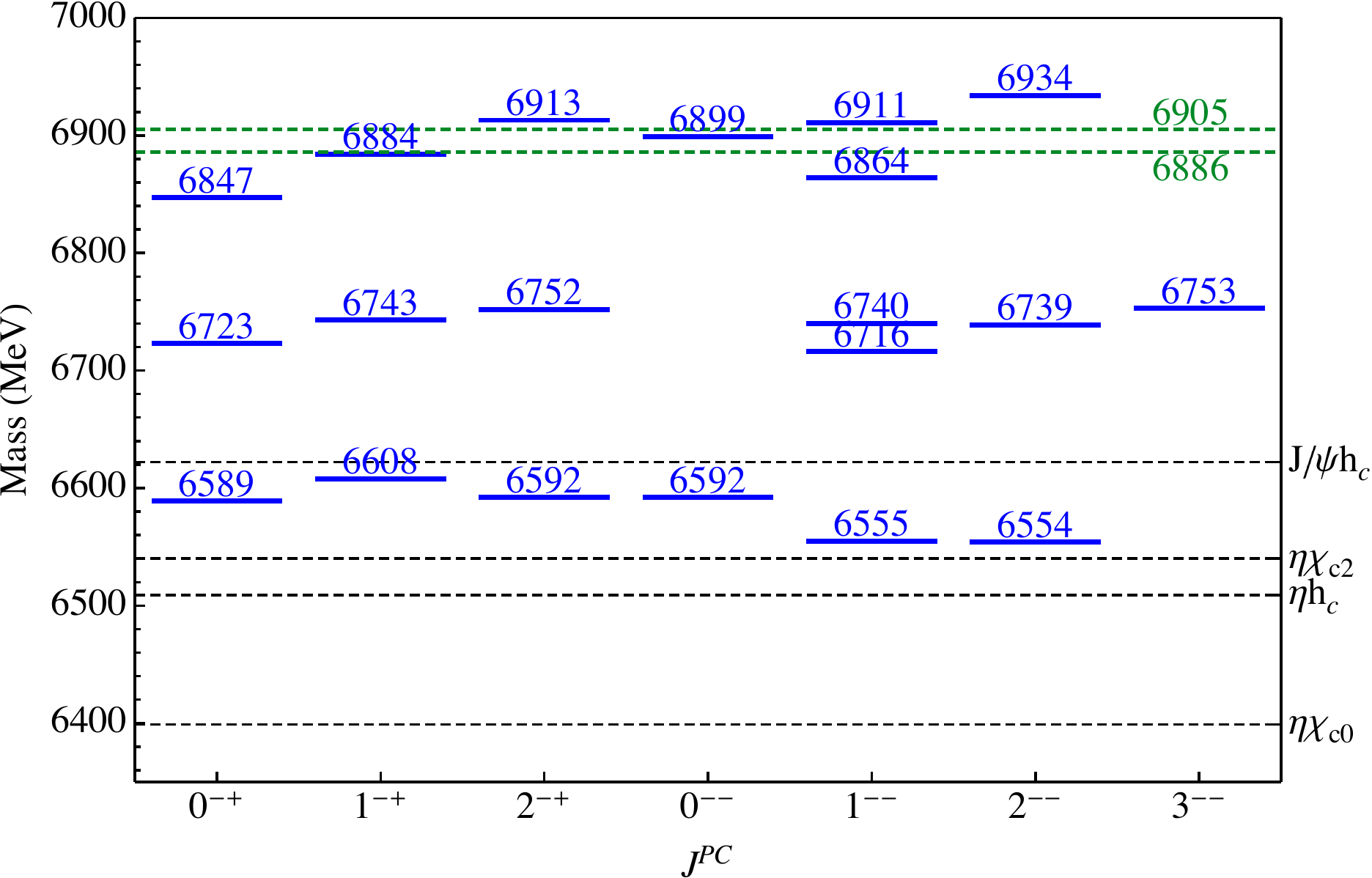}
\caption{The mass spectrum of the P-wave  $T_c$ states. The green
dashed lines stand for the two fitted central mass values of the
$X(6900)$ given by experiments~\cite{Aaij:2020fnh}.}
\label{fig:Pwave}
\end{figure}

For the perturbative potentials, we include the spin-orbital and the
tensor potentials, which contribute to the mixing of the $\lambda$-
and $\rho$-mode states and shift the mass spectrum. We list the
related  spin-orbital and tensor factors in Appendix. We display the
mass spectra of the P-wave states in Fig.~\ref{fig:Pwave} and list
the matrices of the Hamiltonian in Table \ref{tab:pwave}.  The
percentage of the leading eigenstates, and the root mean square
radius are also shown in Table~\ref{tab:pwave}. The root mean square
radii  are smaller than $1$ fm, which indicates that the four quarks
are compactly bound within the tetraquark states.

The spin-orbital and tensor potentials  slightly shift the mass
spectrum obtained with the leading potentials in
Eq.~\eqref{eq:vcen}. The $T_c$ state is dominated by one excitation
mode except the $1^{--}$. The mixing of the two higher $1^{--}$
modes is large, because the $|\lambda_3^-\rangle$ state and the
$|\rho_2^-\rangle$ one are nearly accidentally degenerate after
considering the perturbative mass shifting. Thus, their mixing is
very sensitive to the off-diagonal corrections arising from the
spin-orbital and tensor potentials. For the other P-wave $T_c$'s,
the highest and lightest states are almost totally dominated by
$|\rho_{1}^{+/-}\rangle$ and  $|\rho^{+/-}_{2}\rangle$ components,
thus, are dominated by the  $6_c-\bar 6_c$  and $\bar 3_c-3_c$ color configurations,
respectively.  With the $6_c-\bar 6_c$ components, we obtain a quite
low-lying $0^{--}$ state at $6592$ MeV. Otherwise, the $0^{--}$
state should be located at least $300$ MeV higher. At the same time,
the lowest P-wave state of the other $T_c$'s shall almost be a
$\lambda$-mode excitation.  The decay patterns of the lowest states,
which are sensitive to the $\lambda$- and $\rho$- modes, will be
quite different. Thus, one may use the decay patterns of the $T_c$
states and the mass spectrum of the $0^{--}$ one to study the
$6_c-\bar 6_c$ color configurations.

An unavoidable problem of the framework is that the eigenstates are
much richer than the experimental data. The reasons are listed as
follows. First, in the calculation, we have to solve the
Schr\"odinger equation with finite numbers of basis in the
calculation, which indicates a finite volume of space. Some of the
eigenstates might be the discretized scattering states instead of a
real resonance.  Second, some resonances have large decay widths
and their signals might be overwhelmed by the background  in some
specific channels and are hard to be observed in experiments. What
is more, a broad structure, for instance, the one ranging from
$6.2-6.8$ GeV in LHCb may be a broad resonance or sometimes a
mixture  of several states. To exclude the redundant states and
understand the inner structures of the resonances in experiments, we
need to  corroborate the complex scaling method and calculate decay
widths in the future.

\begin{table*}
 \renewcommand\arraystretch{1.5}
  \caption{The mass spectrum (MeV), the percentages of different $\lambda$- and $\rho$-mode components, and the root mean square radius (fm) of the P-wave tetraquark states. In the second row, we display the mass spectrum  obtained with the leading potentials in Eq.~\eqref{eq:vcen} and the mass corrections from the perturbative spin-orbital and tensor interactions in the mass matrix.}\label{tab:pwave}
 \centering
 \setlength{\tabcolsep}{0.5mm}
\begin{tabular}{c|c|c|ccccc|cccc}
\toprule[1pt] $J^{PC}$&  & Mass & $|\lambda_{1}^{+/-}\rangle$ &
$|\lambda_{2}^{-}\rangle$ & $|\lambda_{3}^{-}\rangle$ &
$|\rho_{1}^{+/-}\rangle$ & $|\rho_{2}^{+/-}\rangle$ &
$r_{12}/r_{34}$ & $r$ & $r_{13}/r_{24}$ & $r'$\tabularnewline \hline
\multirow{3}{*}{$0^{-+}$} &
\multirow{3}{*}{$\left(\begin{array}{ccc}
6746-20 & -20 & -34\\
-20 & 6599+2 & -42\\
-34 & -42 & 6894-62
\end{array}\right)$} & $6589$ & $3.5\%$ &  &  & $92.8\%$ & $3.7\%$ & $0.62$ & $0.33$ & $0.60$ & $0.50$\tabularnewline

 &  & $6723$ & $90.4\%$ &  &  & $5.2\%$ & $4.4\%$ & $0.52$ & $0.43$ & $0.66$ & $0.37$\tabularnewline

 &  & $6847$ & $6.0\%$ &  &  & $2.1\%$ & $91.9\%$ & $0.57$ & $0.38$ & $0.61$ & $0.47$\tabularnewline
\hline \multirow{2}{*}{$0^{--}$} &
\multirow{2}{*}{$\left(\begin{array}{cc}
6561+31 & -11\\
-11 & 6913-14\end{array}\right)$} & $6592$ &  &  &  & $99.9\%$ &
$0.1\%$ & $0.61$ & $0.32$ & $0.59$ & $0.49$\tabularnewline

 &  & $6899$ &  &  &  & $0.1\%$ & $99.9\%$ & $0.58$ & $0.38$ & $0.61$ & $0.48$\tabularnewline
\hline \multirow{3}{*}{$1^{-+}$} &
\multirow{3}{*}{$\left(\begin{array}{ccc}
6746-3 & -4 & -6\\
-4 & 6599+9 & 8\\
-6 & 8 & 6894-10
\end{array}\right)$} & $6608$ & $0.1\%$ &  &  & $99.8\%$ & $0.1\%$ & $0.63$ & $0.33$ & $0.60$ & $0.50$\tabularnewline

 &  & $6743$ & $99.7\%$ &  &  & $0.1\%$ & $0.2\%$ & $0.51$ & $0.43$ & $0.66$ & $0.36$\tabularnewline

 &  & $6884$ & $0.2\%$ &  &  & $0.1\%$ & $99.7\%$ & $0.57$ & $0.37$ & $0.60$ & $0.47$\tabularnewline
\hline \multicolumn{1}{c|}{\multirow{5}{*}{$1^{--}$}} &
\multirow{5}{*}{$\left(\begin{array}{ccccc}
6740 & -2 & 0 & -10 & 9\\
-2 & 6741-23 & 7 & -19 & 26\\
0 & 7 & 6885 & -2 & -25\\
-10 & -19 & -2 & 6561-1 & 21\\
9 & 26 & -25 & 21 & 6913-28
\end{array}\right)$} & $6555$ & $0.3\%$ & $1.6\%$ & $\sim0\%$ & $97.5\%$ & $0.6\%$ & $0.61$ & $0.32$ & $0.59$ & $0.49$\tabularnewline

 &  & $6716$ & $0.6\%$ & $94.8\%$ & $0.4\%$ & $2.0\%$ & $2.1\%$ & $0.52$ & $0.42$ & $0.65$ & $0.37$\tabularnewline

 &  & $6740$ & $98.8\%$ & $0.9\%$ & $\sim0\%$ & $0.2\%$ & $0.1\%$ & $0.51$ & $0.43$ & $0.65$ & $0.36$\tabularnewline

 &  & $6864$ & $0.2\%$ & $2.2\%$ & $55.7\%$ & $0.1\%$ & $41.9\%$ & $0.62$ & $0.35$ & $0.64$ & $0.47$\tabularnewline

 &  & $6911$ & $0.1\%$ & $0.5\%$ & $43.9\%$ & $0.2\%$ & $55.3\%$ & $0.61$ & $0.36$ & $0.63$ & $0.47$\tabularnewline
\hline \multirow{3}{*}{$2^{-+}$} &
\multirow{3}{*}{$\left(\begin{array}{ccc}
6746+6 & 7 & 10\\
7 & 6599-6 & 13\\
10 & 13 & 6894+18
\end{array}\right)$} & $6592$ & $0.2\%$ &  &  & $99.7\%$ & $0.1\%$ & $0.63$ & $0.33$ & $0.60$ & $0.50$\tabularnewline

 &  & $6752$ & $99.4\%$ &  &  & $0.2\%$ & $0.4\%$ & $0.52$ & $0.43$ & $0.66$ & $0.36$\tabularnewline

 &  & $6913$ & $0.4\%$ &  &  & $0.2\%$ & $99.4\%$ & $0.57$ & $0.38$ & $0.60$ & $0.47$\tabularnewline
\hline \multirow{3}{*}{$2^{--}$} &
\multirow{3}{*}{$\left(\begin{array}{ccc}
6741-2 & 7 & -9\\
7 & 6561-6 & -15\\
-9 & -15 & 6913+20
\end{array}\right)$} & $6554$ &  & $0.1\%$ &  & $99.7\%$ & $0.2\%$ & $0.61$ & $0.32$ & $0.59$ & $0.49$\tabularnewline

 &  & $6739$ &  & $99.6\%$ &  & $0.1\%$ & $0.2\%$ & $0.51$ & $0.43$ & $0.66$ & $0.36$\tabularnewline

 &  & $6934$ &  & $0.2\%$ &  & $0.2\%$ & $99.6\%$ & $0.57$ & $0.38$ & $0.61$ & $0.48$\tabularnewline
\hline $3^{--}$ & $6741+11$ & $6753$ &  & $100\%$ &  &  &  & $0.51$
& $0.43$ & $0.66$ & $0.36$\tabularnewline \bottomrule[1pt]
\end{tabular}
\end{table*}

\section{Possible decay modes}\label{sec:decay}

For a $T_c$ state, its decay modes include $T_c \rightarrow  (c\bar
c)+(c\bar c)$, $D^{(*)}\bar D^{(*)}$, light hadrons or $\gamma\gamma$, etc. It
locates above the threshold of two charmonia,  the fall-apart decays
$T_c \rightarrow  (c\bar c)+(c\bar c)$ are therefore dominant. We
have listed some possible low-lying two-charmonium channels in
Table~\ref{tab:decaymode}.

The S-wave radially excited states are much higher than the
low-lying di-$\eta_c$, $J/\psi \eta_c$, or di-$J/\psi$ channels.
They will decay into these channels  with a large phase space and
might be broad. The P-wave $T_c$ states will decay into the lowest
two-charmonium channels di-$\eta_c$, $\eta_c J/\psi$, or di-$
J/\psi$ with a P-wave orbital excitation. These decays are
kinematically suppressed at $\mathcal O (p^3)$ ($p$ is the relative
momentum in the final state). Their S-wave decays modes are the
two-charmonium channels composed of a P-wave charmonium and an
S-wave one. For the low-lying P-wave $T_c$ states, the S-wave decay
modes are suppressed or even not allowed due to the small phase
space. Thus, they may be narrow. For the $0^{--}$ and $1^{-+}$ $T_c$
states, their lowest possible decay modes are the P-wave $\eta_c
J/\psi$ channel and di-$J/\psi$ channel, respectively. These exotic
$J^{PC}$ quantum numbers are not allowed in the conventional
charmonium sector. The search of the above channels will help to
enrich the hadronic spectrum.

 \begin{table*}
    \renewcommand\arraystretch{1.8}
            \caption{The possible decay modes of the $T_c$ states. We use the bold font to emphasize the $J/\psi J/\psi$ mode, where the $X(6900)$ was observed~\cite{Aaij:2020fnh}. }
    \label{tab:decaymode}
    \centering
    \setlength{\tabcolsep}{2.3mm}
    \begin{tabular}{c|c}
        \toprule[1pt]
        $J^{PC}$ & Decay modes \tabularnewline
        \hline
        $0^{++}$ & $\eta_{c}\eta_{c}$, $\bm{J/\psi J/\psi}$, $\chi_{c1}\eta_{c}$(P-wave), $J/\psi h_{c}(1P)$(P-wave), $J/\psi$$\psi(2S)$, $\chi_{c0}\chi_{c0}$\tabularnewline

        $1^{+-}$ & $\eta_{c}J/\psi$, $h_{c}\eta_{c}$(P-wave), $J/\psi\chi_{c1}$(P-wave), $\eta_{c}\psi'$, $h_{c}\chi_{c0}$\tabularnewline

        $2^{++}$ & $\bm{J/\psi J/\psi}$, $\eta_{c}\chi_{c1}$(P-wave), $\eta_{c}\chi_{c2}$(P-wave), $J/\psi h_{c}$(P-wave), $J/\psi \psi(2S)$, $\chi_{c0}\chi_{c2}$\tabularnewline

        $0^{-+}$ & $\bm{J/\psi J/\psi}$(P-wave), $\eta_{c}\chi_{c0}$, $J/\psi h_{c}$, $J/\psi\psi(2S)$(P-wave)\tabularnewline

        $1^{-+}$ &  $\bm{J/\psi J/\psi}$(P-wave) $J/\psi h_{c}$, $J/\psi\psi(2S)$(P-wave)\tabularnewline

        $2^{-+}$ & $\bm{J/\psi J/\psi}$(P-wave), $\eta_{c}\chi_{c2}$, $J/\psi h_{c}$, $J/\psi\psi(2S)$(P-wave)\tabularnewline

        $0^{--}$ & $\eta_{c}J/\psi$(P-wave), $J/\psi\chi_{c1}$, $\eta_{c}\psi(2S)$(P-wave)\tabularnewline

        $1^{--}$ & $\eta_{c}J/\psi$(P-wave), $\eta_{c}h_{c}$, $J/\psi\chi_{c0}$, $J/\psi\chi_{c1}$, $J/\psi\chi_{c2}$, $\eta_{c}\psi'$(P-wave)\tabularnewline

        $2^{--}$ & $\eta_{c}J/\psi$(P-wave), $J/\psi\chi_{c1}$, $J/\psi\chi_{c2}$, $\eta_{c}\psi'$(P-wave), $h_{c}\chi_{c0}$(P-wave)\tabularnewline

        $3^{--}$ & $J/\psi\chi_{c2}$\tabularnewline
        \bottomrule[1pt]
    \end{tabular}
 \end{table*}

For the di-$J/\psi$  decay mode, the possible $J^{PC}$ quantum
numbers are  $0^{++}$ and  $2^{++}$ for the S-wave states, while
those are $0^{-+}$, $1^{-+}$, and  $2^{-+}$ for the P-wave states.
Based on the mass spectra, the first radial excitations with the
masses $M(0^{++})=6867$ MeV and $M(2^{++})=6917$ MeV, or the
$1^{-+}$ and $2^{-+}$ P-wave states with the masses $M(1^{-+})=6884$
MeV and $M=6913$ MeV, may be the candidates for the $X(6900)$. As an
S-wave radial excitation, it will decay into the S-wave di-$\eta_c$
or di-$J/\psi$ channels. As a P-wave $T_c$ state, its decays into
the two final states are kinetically suppressed. The other S-wave
decay modes are $J/\psi h_{c}$ for the $1^{-+}$ $T_c$ state,
$\eta_c\chi_{c2}$ and $J/\psi h_c$ for the $2^{-+}$ one with smaller
phase space. Thus, the $X(6900)$ as an S-wave $T_c$ state should be
much broader. We can determine the $J^{PC}$ quantum numbers of the
$X(6900)$ by studying  their decay patterns in the future.

\section{Discussions}\label{sec3}
In Sec.~\ref{sec:pwv}, we obtain the relation among  the P-wave
$\rho$-mode and $\lambda$-mode excitations with different color
configurations, $6_{\rho}<3_{\lambda}<6_{\lambda}<3_{\rho}$.
Moreover, if we compare the mass spectra of the S-wave and P-wave
$T_c$ states as illustrated in Fig.~\ref{fig:Swave} and
Fig.~\ref{fig:Pwave},  the mass gaps are much smaller than those of
the charmonia. In this section, we will give a qualitative
discussion about the two results.

In the baryon sector, if the confinement potential dominates the
binding, one can use a phenomenological harmonic oscillator
potential to discuss the feature of the two excitation modes for the
P-wave baryons~\cite{Yoshida:2015tia}. We expand the method to
discuss the $T_c$ state.

For the $3_c-\bar 3_c$ confined $T_c$ state, the Hamiltonian in the
harmonic oscillator potential is written as
\begin{eqnarray}
&&H =\sum_{i=1}^{4}\frac{\mathbf{p}_{i}^{2}}{2m_{i}}+\frac{k}{2}(r_{12}^{2}+r_{34}^{2})+\frac{k'}{4}(r_{13}^{2}+r_{24}^{2}+r_{14}^{2}+r_{23}^{2})\nonumber\\
&&  =\frac{\mathbf{p}_{a}^{2}}{2u_{a}}+\frac{\mathbf{p}_{b}^{2}}{2u_{b}}+\frac{\mathbf{p}_{ab}^{2}}{2u_{ab}}+\frac{u_{a}\omega_{a}^{2}}{2}r_{12}^{2}+\frac{u_{b}\omega_{b}^{2}}{2}r_{34}^{2}+\frac{u_{ab}\omega_{ab}^{2}}{2}r^{2},\nonumber\\
\end{eqnarray}
where $\mathbf{p}_{a/b/ab}$ and $u_{a/b/ab}$ are the relative
momenta and the reduced masses of the diquark/antidiquarks, or those
between the two clusters. $k$ and $k'$ are the coefficients of the
confinement potential. The frequencies $\omega$ are given by
\begin{eqnarray}\label{rhofre}
\omega_{a}=\sqrt{\frac{2k+k'}{2u_{a}}},\,\omega_{b}=\sqrt{\frac{2k+k'}{2u_{b}}},\,\omega_{ab}=\sqrt{\frac{2k'}{u_{ab}}}.
\end{eqnarray}
The eigenvalue then reads
\begin{eqnarray}
E&=&(2n_{a}+l_{a}+3/2)\hbar\omega_{a}+(2n_{b}+l_{b}+3/2)\hbar\omega_{b}\nonumber\\
&+&(2n_{ab}+l_{ab}+3/2)\hbar\omega_{ab}.
\end{eqnarray}
For the fully heavy tetraquark state $QQ\bar Q\bar Q$, one has
$u_a=u_b=\frac{m_{Q}}{2}$ and $u_{ab}=m_Q$. The coupling constant
$k$ and $k'$ should be similar to each other according to the
experiences in the conventional hadrons. If we assume $k=k'$, one
has
\begin{eqnarray}\label{rhofre}
\omega_{a}=\omega_{b}=\sqrt{\frac{3k}{m_Q}}>\,\omega_{ab}=\sqrt{\frac{2k}{m_Q}},
\end{eqnarray}
which indicates that the $ 3_{\lambda}$ mode P-wave state is located
lower than the $ 3_\rho$ mode ones. This  is also a feature of the
P-wave baryons~\cite{Yoshida:2015tia}.

For the $6_c-\bar 6_c$ state, we can write the Hamiltonian
similarly,
\begin{eqnarray}
&&H =\sum_{i}\frac{p_{i}^{2}}{2m_{i}}-\frac{k}{4}(r_{12}^{2}+r_{34}^{2})+\frac{5k'}{8}(r_{13}^{2}+r_{24}^{2}+r_{14}^{2}+r_{23}^{2})\nonumber\\
&&  =\frac{p_{a}^{2}}{2u_{a}}+\frac{p_{b}^{2}}{2u_{b}}+\frac{p_{ab}^{2}}{2u_{ab}}+\frac{u_{a}\omega_{a}^{2}}{2}r_{12}^{2}+\frac{u_{b}\omega_{b}^{2}}{2}r_{34}^{2}+\frac{u_{ab}\omega_{ab}^{2}}{2}r^{2},\nonumber\\
\end{eqnarray}
with the frequencies as
\begin{eqnarray}\label{lambdafre}
    \omega_{a}=\sqrt{\frac{-2k+5k'}{4u_{a}}},\,\omega_{b}=\sqrt{\frac{-2k+5k'}{4u_{b}}},\,\omega_{ab}=\sqrt{\frac{5k'}{u_{ab}}}. \nonumber\\
\end{eqnarray}
If $k=k'$, one obtains
\begin{eqnarray}\label{lambdafre}
    \omega_{a}=\omega_{b}=\sqrt{\frac{3k}{2m_Q}}<\,\omega_{ab}=\sqrt{\frac{5k}{m_Q}},
    \end{eqnarray}
which indicates  that the $6_{\lambda}$ mode state is located higher
than the $6_{\rho}$ mode one.

With Eq. (\ref{rhofre}) and Eq. (\ref{lambdafre}),  we obtain  the
qualitative relationship among the mass spectra
\begin{eqnarray}
6_\rho<3_\lambda<3_\rho<6_\lambda.
\end{eqnarray}
This is consistent with the relation in Table~\ref{tab:lambdarholo}
except that the $3_\rho$ is lower than the $6_\lambda$, where the
relation is inverse in the dynamical calculations.

We can also write the Hamiltonian of the heavy quarkonium in the
harmonic oscillator potential as follows:
\begin{eqnarray}
&&H =\sum_{i}\frac{p_{i}^{2}}{2m_{i}}+k
r_{12}^{2}=\frac{p^{2}}{2u_m}+\frac{u_m\omega^{2}}{2}r_{12}^{2},
\end{eqnarray}
with
\begin{eqnarray}
&&u_m=\frac{m_Q}{2}, ~~~\omega_m=\sqrt{\frac{4k}{m_Q}},
\end{eqnarray}
where $u_m$ and $\omega_m$ are the reduced mass and the frequency of
the charmonium, respectively. In general, the mass splitting for the
ground S-wave and P-wave charmonium  $\hbar\sqrt{\frac{4k}{m_c}}$ is
about $400\sim 500$ MeV. In contrast, the mass splitting for the
ground S-wave and P-wave $T_c$ state is
$\hbar\sqrt{\frac{3k}{2m_c}}$ and its value is about $245\sim 300$
MeV. Thus, the small mass difference for the S-wave and P-wave $T_c$
state arises from the low-lying $6_\rho$ component.

\section{Summary} \label{sec4}

In this work, we have systematically calculated the mass spectra of
the S-wave and P-wave  $T_c$ states and discussed their structures
with a  nonrelativistic quark model. The parameters are determined
by the spectrum of the  charmonium. In the calculation, we treat the
OGE Coulomb-like, linear confinement, and hyperfine interactions as
leading order potentials. The remaining spin-orbital and tensor
potentials are treated as the perturbative interactions, which will
contribute to the mass shifts. 

In the analysis of the $T_c$  wave
function, we include both the $\bar 3_c-3_c$ and the $6_c-\bar 6_c$
color configurations. To obtain the mass spectrum, we solve the
Schr\"odinger equation with the variational method. For the S-wave $T_c$ states, we do not consider the perturbative  potentials since the induced mass shifts should be much smaller than the mass differences  arising from the leading potentials and will not influence  the mass spectrum significantly.  For the P-wave state, the spin-orbital and tensor potentials are essential to distinguish the $T_c$ states with different $J^{PC}$ quantum numbers.  Their   contributions are calculated  in the basis of the eigenvectors which are obtained  with only the leading  potentials.  We give the
numerical results for the S-wave $T_c$ states plus their radial
excitations, and the P-wave $T_c$ states in Fig.~\ref{fig:Swave} and
Fig.~\ref{fig:Pwave}, respectively.

For the S-wave state, we find that the ground state is located above
the low-lying two-charmonium channels di- $\eta_c$, $J/\psi \eta_c$
and di-$J/\psi$. The $6_c-\bar 6_c$  color configuration only
appears in the $0^{++}$ state and dominates the ground state as
illustrated in Table~\ref{tab:swave}. In other words, the $6_c-\bar
6_c$ component is located lower than the $\bar 3_c-3_c$ one, because
the prior one has an attractive and much stronger interactions
between the diquark and antidiquark. The fall-apart decays of the
S-wave $T_c$ states into two charmonia imply that these states,
especially their radial excitations might be very broad. For the
$X(6900)$ state, the ground states are located much lower, while the
first radial excitation of the $0^{++}$ and $2^{++}$ state lie very
close. If it turns out to be the first radial excitation, it shall
have a broad decay width.

For the P-wave state, we focus on two excited modes, i.e,. the
$\lambda$-mode and $\rho$-mode, and their properties. The two
excitation modes with the negative charge parity do not couple with
each other, while the modes with the positive C-parity mix quite
slightly because of the hyperfine potential. The relation among
different modes with different color configurations is
$6_{\rho}<3_{\lambda}<6_{\lambda}<3_{\rho}$.  At the next order, the
spin-orbital and tensor potentials cause the small mass shifting and
their mixing. Almost all of the mixing is small and one mode is
dominant as illustrated in Table~\ref{tab:pwave}. The mass spectrum
of the P-wave $\rho$-mode $T_c$  state is quite sensitive to the
color configurations, which is useful to investigate the role of the
rare $6_c-\bar 6_c$ color configuration in the multiquark system.
Moreover, the small mass gap between the P-wave  and the ground
S-wave $T_c$ states is due to low-lying $\rho$-mode excitations with
the $6_c-\bar 6_c$ color configuration.   The masses of the $1^{-+}$
and $2^{-+}$ tetraquarks are $6884$ MeV and $6913$ MeV,
respectively, which is consistent with the mass of the $X(6900)$
state. If we treat it as a P-wave $T_c$ state, as illustrated in
Table~\ref{tab:decaymode}, its decays into the lowing state are
kinematically suppressed. Its S-wave decay modes contain an excited
P-wave charmonium and have a small phase space. Thus, the decay
width of the P-wave tetraquarks are expected to be narrower than the
S-wave ones. More studies of the decay patterns will be useful to
determine its $J^{PC}$ quantum numbers. Besides $X(6900)$, there may
exist many other tetraquark states. The $T_c$ states with the $J^{PC}=0^{--}$ and $1^{-+}$ are of special interest since they are not allowed in the conventional  quark model. The  lowest states are narrow and the phase-space-allowed  decay modes are the P-wave $\eta_c J/\psi $ and di-$J/\psi$ channels, respectively. If observed, they are unambiguously the multiquark system, which will enrich the hadron spectroscopy.  Hopefully, our calculations will
be useful for the search of the new exotic tetraquark states. More
experimental and theoretical works are expected to verify and
understand the multiquark system in the future.

\appendix
\section{The spin, spin-orbital and tensor factors }

In this section, we present the related color magnetic factors in
Table \ref{tab:csfactor}. To calculate the spin factors, we
extract the spin wave functions of different
color-flavor configurations listed in Table \ref{tab:csfwf}.
 For the $J^{PC}=0^{-+}$, $1^{-+}$, $2^{-+}$ states, the spin wave function of the $\lambda$-mode is
\begin{eqnarray}
\chi^{\lambda}_{s}=\left[[QQ]_{\bar{3}_{c}}^{1}[\bar{Q}_{2}\bar{Q}_{3}]_{3_{c}}^{1}\right]_{1_{c}}^{1}.
\end{eqnarray}
The spin wave function of the two $\rho$-mode excitations  read,
\begin{eqnarray}
&&{\left[\left[[QQ]_{\bar{3}_{c}}^{0},\text{\ensuremath{\rho}}\right]_{\bar{3}_{c}}^{1}[\bar {Q}\bar {Q}]_{3_{c}}^{1}\right]^{0}}:~\chi^{\rho}_{s1} = \left([QQ]_{\bar{3}_{c}}^{0}[\bar {Q}\bar {Q}]_{3_{c}}^{1}\right)^{1}, \nonumber \\
&&{\left[\left[[QQ]_{6_{c}}^{1},\rho\right]_{6_{c}}^{0}[\overline{Q}\bar{Q}]_{\bar{6}_{c}}^{0}\right]^{0}}:~\chi^{\rho}_{s2}={\left([QQ]_{6_{c}}^{1}[\bar {Q}\bar{Q}]_{\bar{6}_{c}}^{0}\right)^{1}}.\nonumber \\
\end{eqnarray}
Here, we select the spin wave functions for the  the
color-flavor-spin configurations with  the P-wave excitation in the
diquark as an example. The corresponding color magnetic factors are
listed in Table  \ref{tab:csfactor}.  The P-wave may also appear
in the antidiquark. The calculation of their factors is similar.

The spin-orbital plus the tensor factors are listed in Table
\ref{lolambda}- Table \ref{lolambdarho}.

\begin{table*}
 \caption{The color magnetic factors $\langle H_{CM}^{ij}=\frac{\lambda_{i}}{2}\cdot \frac{\lambda_{j}}{2}s_{i}\cdot s_{j}\rangle $ for a fully heavy tetraquark state $Q_1Q_2\bar Q_3\bar Q_4$.  } \label{tab:csfactor}
 \renewcommand\arraystretch{2.0}
 \centering
 \setlength{\tabcolsep}{2.3mm}
\begin{tabular}{ccccccc}
\toprule[1pt]
&$\langle H_{CM}^{Q_{1}\bar{Q}_{3}}\rangle$ & $\langle
H_{CM}^{Q_{2}\bar{Q}_{4}}\rangle$ & $\langle
H_{CM}^{Q_{1}\bar{Q}_{4}}\rangle$ & $\langle
H_{CM}^{Q_{2}\bar{Q}_{3}}\rangle$ & $\langle
H_{CM}^{Q_{1}Q_{2}}\rangle$ & $\langle
H_{CM}^{\bar{Q}_{3}\bar{Q}_{4}}\rangle$\tabularnewline \hline
{$\langle \chi^{\lambda}_s |H^{ij}_{CM}|\chi^{\lambda}_s\rangle$} &
$\frac{1}{12}$ & $\frac{1}{12}$ & $\frac{1}{12}$ & $\frac{1}{12}$ &
$-\frac{1}{6}$ & $-\frac{1}{6}$\tabularnewline

$\langle \chi^{\rho}_{s1}|H^{ij}_{CM}|\chi^{\rho}_{s1}\rangle$  &$0$
& $0$ & $0$ & $0$ & $\frac{1}{2}$ & $-\frac{1}{6}$ \tabularnewline

$\langle \chi^{\rho}_{s2}|H^{ij}_{CM}|\chi^{\rho}_{s2}\rangle$
&$0$ & $0$ & $0$ & $0$ &
$\frac{1}{12}$  & $-\frac{1}{4}$\tabularnewline

$\langle \chi^{\lambda}_{s} |H^{ij}_{CM}| \chi^{\rho}_{s1} \rangle$
& $\frac{1}{6 \sqrt{2}}$ & $-\frac{1}{6 \sqrt{2}}$ & $\frac{1}{6
\sqrt{2}}$ & $-\frac{1}{6 \sqrt{2}}$ & $0$ & $0$\tabularnewline

$\langle \chi^{\lambda}_{s} |H^{ij}_{CM}| \chi^{\rho}_{s2} \rangle$
& $-\frac{1}{4}$ & $\frac{1}{4}$ & $-\frac{1}{4}$ & $\frac{1}{4}$ & $0$ & $0$\tabularnewline

$\langle \chi^{\rho}_{s1}  |H^{ij}_{CM}| \chi^{\rho}_{s2} \rangle$ &
$-\frac{1}{4\sqrt{2}}$ & $-\frac{1}{4\sqrt{2}}$ &
$-\frac{1}{4\sqrt{2}}$ & $-\frac{1}{4\sqrt{2}}$ & $0$ &
$0$\tabularnewline \bottomrule[1pt]
\end{tabular}
\end{table*}

\begin{table*}
 \renewcommand\arraystretch{1.8}
  \caption{The spin-orbital and tensor factors for the $\lambda$-mode P-wave states. }\label{lolambda}
 \centering
 \setlength{\tabcolsep}{0.1mm}

\begin{tabular}{c|c|c|c|cccccc}
\toprule[1pt]

 $J^{PC}$ & &  & $J$ & $Q_{1}Q_{2}$ & $\bar{Q}_{3}\bar{Q}_{4}$ & $Q_{1}\bar{Q}_{3}$ & $Q_{2}\bar{Q}_{4}$ & $Q_{1}\bar{Q}_{4}$ & $Q_{2}\bar{Q}_{3}$ \tabularnewline
\hline \multirow{6}{*}{$J^{-+}$} &
\multirow{6}{*}{$\left[[QQ]_{\bar{3}_{c}}^{1}[\bar{Q}\bar{Q}]_{3_{c}}^{1},\lambda\right]^{J}$}
& \multirow{3}{*}{$\mathbf{L}_{ij}\cdot\mathbf{S}_{ij}$} & $J=0$ &
$0$ & $0$ & $-1$ & $-1$ & $-1$ & $-1$\tabularnewline

 &  &  & $J=1$ & $0$ & $0$ & $-\frac{1}{2}$ & $-\frac{1}{2}$ & $-\frac{1}{2}$ & $-\frac{1}{2}$\tabularnewline

 &  &  & $J=2$ & $0$ & $0$ & $\frac{1}{2}$ & $\frac{1}{2}$ & $\frac{1}{2}$ & $\frac{1}{2}$\tabularnewline
\cline{3-10} &  & \multirow{3}{*}{$\mathcal{S}_{ij}$} & $J=0$ & $0$
& $0$ & $-\frac{1}{2}$ & $-\frac{1}{2}$ & $-\frac{1}{2}$ &
$-\frac{1}{2}$\tabularnewline

 &  &  & $J=1$ & $0$ & $0$ & $\frac{1}{4}$ & $\frac{1}{4}$ & $\frac{1}{4}$ & $\frac{1}{4}$\tabularnewline

 &  &  & $J=2$ & $0$ & $0$ & $-\frac{1}{20}$ & $-\frac{1}{20}$ & $-\frac{1}{20}$ & $-\frac{1}{20}$\tabularnewline
\hline \multirow{12}{*}{$1^{--}$} &
 & \multirow{4}{*}{$\mathbf{L}_{ij}\cdot\mathbf{S}_{ij}$}
 & \multirow{1}{*}{$\langle\phi_{1}({}^{1}P_{1})|\hat{O}|\phi_{1}({}^{1}P_{1})\rangle$} & $0$ & $0$ & $0$ & $0$ & $0$ & $0$\tabularnewline

&  \multirow{4}{*}{$
\phi_{1}(^{1}P_{1})=\left[\left[[QQ]_{\bar{3}_{c}}^{1}[\bar{Q}\bar{Q}]_{3_{c}}^{1}\right]_{1_{c}}^{0},\lambda\right]_{1_{c}}^{1}$}
&  & $\langle \phi_{2}(^{1}P_{1})|\hat{O}|\phi_{2}(^{1}P_{1})
\rangle$ & $0$ & $0$ & $0$ & $0$ & $0$ & $0$\tabularnewline

 & &  & $\langle\phi_{3}({}^{5}P_{1})|\hat{O}|\phi_{3}({}^{5}P_{1})\rangle$ & $0$ & $0$ & $-\frac{3}{2}$ & $-\frac{3}{2}$ & $-\frac{3}{2}$ & $-\frac{3}{2}$\tabularnewline

 &  &  & $\langle\phi_{1}({}^{1}P_{1})|\hat{O}|\phi_{2}({}^{1}P_{1})\rangle$ & $0$ & $0$ & $0$ & $0$ & $0$ & $0$\tabularnewline

 &\multirow{4}{*}{$ \phi_{2}(^{1}P_{1})=\left[\left[[QQ]_{6_{c}}^{0}[\bar{Q}\bar{Q}]_{\bar{6}_{c}}^{0}\right]_{1_{c}}^{0},\lambda\right]_{1_{c}}^{1}$}    &  & $\langle\phi_{1}({}^{1}P_{1})|\hat{O}|\phi_{3}({}^{5}P_{1})\rangle$ & $0$ & $0$ & $0$ & $0$ & $0$ & $0$\tabularnewline

 &  &  & $\langle\phi_{2}({}^{1}P_{1})|\hat{O}|\phi_{3}({}^{5}P_{1})\rangle$ & $0$ & $0$ & $0$ & $0$ & $0$ & $0$\tabularnewline
\cline{3-10}
 &  & \multirow{6}{*}{$\mathcal{S}_{ij}$} & $\langle\phi_{1}({}^{1}P_{1})|\hat{O}|\phi_{1}({}^{1}P_{1})\rangle$ & $0$ & $0$ & $0$ & $0$ & $0$ & $0$\tabularnewline

 &  &  & $\langle\phi_{2}({}^{1}P_{1})|\hat{O}|\phi_{2}({}^{1}P_{1})\rangle$ & $0$ & $0$ & $0$ & $0$ & $0$ & $0$\tabularnewline

 & $\phi_{3}(^{5}P_{1})=\left[\left[[QQ]_{\bar{3}_{c}}^{1}[\bar{Q}\bar{Q}]_{3_{c}}^{1}\right]_{1_{c}}^{2},\lambda\right]_{1_{c}}^{1}$  &  & $\langle\phi_{3}({}^{5}P_{1})|\hat{O}|\phi_{3}({}^{5}P_{1})\rangle$ & $0$ & $0$ & $-\frac{7}{20}$ & $-\frac{7}{20}$ & $-\frac{7}{20}$ & $-\frac{7}{20}$\tabularnewline

 &  &  & $\langle\phi_{1}({}^{1}P_{1})|\hat{O}|\phi_{2}({}^{1}P_{1})\rangle$ & $0$ & $0$ & $0$ & $0$ & $0$ & $0$\tabularnewline

 &  &  & $\langle\phi_{1}({}^{1}P_{1})|\hat{O}|\phi_{3}({}^{5}P_{1})\rangle$ & $0$ & $0$ & $\frac{1}{2\sqrt{5}}$ & $\frac{1}{2\sqrt{5}}$ & $\frac{1}{2\sqrt{5}}$ & $\frac{1}{2\sqrt{5}}$\tabularnewline

 &  &  & $\langle\phi_{2}({}^{1}P_{1})|\hat{O}|\phi_{3}({}^{5}P_{1})\rangle$ & $0$ & $0$ & $-\frac{\sqrt{\frac{3}{5}}}{2}$ & $-\frac{\sqrt{\frac{3}{5}}}{2}$ & $\frac{\sqrt{\frac{3}{5}}}{2}$ & $\frac{\sqrt{\frac{3}{5}}}{2}$\tabularnewline
\hline \multirow{2}{*}{$2^{--}$} & \multirow{2}{*}{$
\left[\left[[QQ]_{\bar{3}_{c}}^{1}[\bar{Q}\bar{Q}]_{3_{c}}^{1}\right]_{1_{c}}^{2},\lambda\right]_{1_{c}}^{2}$}
& $\mathbf{L}_{ij}\cdot\mathbf{S}_{ij}$ &
\multirow{2}{*}{$\langle\phi_{1}({}^{5}P_{2})|\hat{O}|\phi_{1}({}^{5}P_{2})\rangle$}
& $0$ & $0$ & $-\frac{1}{2}$ & $-\frac{1}{2}$ & $-\frac{1}{2}$ &
$-\frac{1}{2}$\tabularnewline

 &  & $\mathcal{S}_{ij}$ &  & $0$ & $0$ & $\frac{7}{20}$ & $\frac{7}{20}$ & $\frac{7}{20}$ & $\frac{7}{20}$\tabularnewline
\hline \multirow{2}{*}{$3^{--}$} &
\multirow{2}{*}{$\left[\left[[QQ]_{\bar{3}_{c}}^{1}[\bar{Q}\bar{Q}]_{3_{c}}^{1}\right]_{1_{c}}^{2},\lambda\right]_{1_{c}}^{3}$}
& $\mathbf{L}_{ij}\cdot\mathbf{S}_{ij}$ &
\multirow{2}{*}{$\langle\phi_{1}({}^{5}P_{3})|\hat{O}|\phi_{1}({}^{5}P_{3})\rangle$}
& $0$ & $0$ & $1$ & $1$ & $1$ & $1$\tabularnewline

 &  & $\mathcal{S}_{ij}$ &  & $0$ & $0$ & $-\frac{1}{10}$ & $-\frac{1}{10}$ & $-\frac{1}{10}$ & $-\frac{1}{10}$\tabularnewline
\hline
\end{tabular}
\end{table*}

\begin{table*}
 \renewcommand\arraystretch{2.1}
  \caption{The spin-orbital and tensor factors of the $\rho$-mode P-wave states.  The position of the matrix elements corresponds to the subscripts of the color-flavor-spin configurations. }\label{lorho}
 \centering
 \setlength{\tabcolsep}{0.8mm}
    \resizebox{\textwidth}{11.5cm}{
\begin{tabular}{c|c|cccc}
\toprule[1pt] \multicolumn{1}{c|}{$J^{P}$} &  &
\multicolumn{4}{c}{$\mathbf{L}_{ij}\cdot\mathbf{S}_{ij}$}
\tabularnewline \cline{3-6} \multicolumn{1}{c|}{} &  & $Q_{1}Q_{2}$
& $\bar{Q}_{3}\bar{Q}_{4}$ & $Q_{1}\bar{Q}_{3}$/$Q_{2}\bar{Q}_{4}$ &
$Q_{1}\bar{Q}_{4}$/$Q_{2}\bar{Q}_{3}$\tabularnewline \hline
\multicolumn{1}{c|}{\multirow{4}{*}{$0^{-}$}} &
$\phi_{1}=\left[\left[[QQ]_{\bar{3}_{c}}^{0},\text{\ensuremath{\rho}}\right]_{\bar{3}_{c}}^{1}[\bar{Q}\bar{Q}]_{3_{c}}^{1}\right]^{0}$
& \multirow{4}{*}{$\left(\begin{array}{cccc}
0 & 0 & 0 & 0\\
0 & 0 & 0 & 0\\
0 & 0 & -2 & 0\\
0 & 0 & 0 & 0
\end{array}\right)$} & \multirow{4}{*}{$\left(\begin{array}{cccc}
0 & 0 & 0 & 0\\
0 & 0 & 0 & 0\\
0 & 0 & 0 & 0\\
0 & 0 & 0 & -2
\end{array}\right)$} & \multirow{4}{*}{$\left(\begin{array}{cccc}
-1 & 0 & 0 & -1\\
0 & -1 & -1 & 0\\
0 & -1 & -1 & 0\\
-1 & 0 & 0 & -1
\end{array}\right)$} & \multirow{4}{*}{$\left(\begin{array}{cccc}
-1 & 0 & 0 & -1\\
0 & -1 & -1 & 0\\
0 & -1 & -1 & 0\\
-1 & 0 & 0 & -1
\end{array}\right)$}\tabularnewline
 & $\phi_{2}=\left[[QQ]_{\bar{3}_{c}}^{1}\left[[\bar{Q}\bar{Q}]_{3_{c}}^{0},\text{\ensuremath{\rho}}\right]_{\bar{3}_{c}}^{1}\right]^{0}$ &  &  &  & \tabularnewline
 & $\phi_{3}=\left[\left[[QQ]_{6_{c}}^{1},\rho\right]_{6_{c}}^{0}[\bar{Q}\bar{Q}]_{\bar{6}_{c}}^{0}\right]^{0}$ &  &  &  & \tabularnewline
 & $\phi_{4}=\left[\left[QQ\right]_{6_{c}}^{0}\left[[\bar{Q}\bar{Q}]_{\bar{6}_{c}}^{1},\rho\right]_{\bar{6}_{c}}^{0}\right]^{0}$ &  &  &  & \tabularnewline
\hline \multirow{4}{*}{$1^{-}$} &
$\phi_{1}=\left[\left[[QQ]_{\bar{3}_{c}}^{0},\text{\ensuremath{\rho}}\right]_{\bar{3}_{c}}^{1}[\bar{Q}\bar{Q}]_{3_{c}}^{1}\right]^{1}$
& \multirow{4}{*}{$\left(\begin{array}{cccc}
0 & 0 & 0 & 0\\
0 & 0 & 0 & 0\\
0 & 0 & -1 & 0\\
0 & 0 & 0 & 0
\end{array}\right)$} & \multirow{4}{*}{$\left(\begin{array}{cccc}
0 & 0 & 0 & 0\\
0 & 0 & 0 & 0\\
0 & 0 & 0 & 0\\
0 & 0 & 0 & -1
\end{array}\right)$} & \multirow{4}{*}{$\left(\begin{array}{cccc}
-\frac{1}{2} & 0 & 0 & -\frac{1}{2}\\
0 & -\frac{1}{2} & -\frac{1}{2} & 0\\
0 & -\frac{1}{2} & -\frac{1}{2} & 0\\
-\frac{1}{2} & 0 & 0 & -\frac{1}{2}
\end{array}\right)$} & \multirow{4}{*}{$\left(\begin{array}{cccc}
-\frac{1}{2} & 0 & 0 & -\frac{1}{2}\\
0 & -\frac{1}{2} & -\frac{1}{2} & 0\\
0 & -\frac{1}{2} & -\frac{1}{2} & 0\\
-\frac{1}{2} & 0 & 0 & -\frac{1}{2}
\end{array}\right)$}\tabularnewline

 & $\phi_{2}=\left[[QQ]_{\bar{3}_{c}}^{1}\left[[\bar{Q}\bar{Q}]_{3_{c}}^{0},\text{\ensuremath{\rho}}\right]_{\bar{3}_{c}}^{1}\right]^{1}$ &  &  &  & \tabularnewline

 & $\phi_{3}=\left[\left[[QQ]_{6_{c}}^{1},\rho\right]_{6_{c}}^{1}[\bar{Q}\bar{Q}]_{\bar{6}_{c}}^{0}\right]^{1}$ &  &  &  & \tabularnewline

 & $\phi_{4}=\left[\left[QQ\right]_{6_{c}}^{0}\left[[\bar{Q}\bar{Q}]_{\bar{6}_{c}}^{1},\rho\right]_{\bar{6}_{c}}^{1}\right]^{1}$ &  &  &  & \tabularnewline
\hline \multicolumn{1}{c|}{\multirow{4}{*}{$2^{-}$}} &
$[\phi_{1}=\left[\left[[QQ]_{\bar{3}_{c}}^{0},\text{\ensuremath{\rho}}\right]_{\bar{3}_{c}}^{1}[\bar{Q}\bar{Q}]_{3_{c}}^{1}\right]^{2}]$
 & \multirow{4}{*}{$\left(\begin{array}{cccc}
0 & 0 & 0 & 0\\
0 & 0 & 0 & 0\\
0 & 0 & 1 & 0\\
0 & 0 & 0 & 0
\end{array}\right)$} & \multirow{4}{*}{$\left(\begin{array}{cccc}
0 & 0 & 0 & 0\\
0 & 0 & 0 & 0\\
0 & 0 & 0 & 0\\
0 & 0 & 0 & 1
\end{array}\right)$} & \multirow{4}{*}{$\left(\begin{array}{cccc}
\frac{1}{2} & 0 & 0 & \frac{1}{2}\\
0 & \frac{1}{2} & \frac{1}{2} & 0\\
0 & \frac{1}{2} & \frac{1}{2} & 0\\
\frac{1}{2} & 0 & 0 & \frac{1}{2}
\end{array}\right)$} & \multirow{4}{*}{$\left(\begin{array}{cccc}
\frac{1}{2} & 0 & 0 & \frac{1}{2}\\
0 & \frac{1}{2} & \frac{1}{2} & 0\\
0 & \frac{1}{2} & \frac{1}{2} & 0\\
\frac{1}{2} & 0 & 0 & \frac{1}{2}
\end{array}\right)$}\tabularnewline
  & $\phi_{2}=\left[[QQ]_{\bar{3}_{c}}^{1}\left[[\bar{Q}\bar{Q}]_{3_{c}}^{0},\text{\ensuremath{\rho}}\right]_{\bar{3}_{c}}^{1}\right]^{2}$ &  &  &  & \tabularnewline

 & $\phi_{3}=\left[\left[[QQ]_{6_{c}}^{1},\rho\right]_{6_{c}}^{2}[\bar{Q}\bar{Q}]_{\bar{6}_{c}}^{0}\right]^{2}$ &  &  &  & \tabularnewline

 & $\phi_{4}=\left[\left[QQ\right]_{6_{c}}^{0}\left[[\bar{Q}\bar{Q}]_{\bar{6}_{c}}^{1},\rho\right]_{\bar{6}_{c}}^{2}\right]^{2}$ &  &  &  & \tabularnewline
\hline
 & & \multicolumn{4}{c}{$\mathcal{S}_{ij}$}\tabularnewline
\hline

 \multirow{4}{*}{$0^{-}$} & \multicolumn{1}{c}{$\phi_{1}=\left[\left[[QQ]_{\bar{3}_{c}}^{0},\text{\ensuremath{\rho}}\right]_{\bar{3}_{c}}^{1}[\bar{Q}\bar{Q}]_{3_{c}}^{1}\right]^{0}$}  & \multicolumn{1}{c}{\multirow{4}{*}{$\left(\begin{array}{cccc}
0 & 0 & 0 & 0\\
0 & 0 & 0 & 0\\
0 & 0 & -1 & 0\\
0 & 0 & 0 & 0
\end{array}\right)$}} & \multicolumn{1}{c}{\multirow{4}{*}{$\left(\begin{array}{cccc}
0 & 0 & 0 & 0\\
0 & 0 & 0 & 0\\
0 & 0 & 0 & 0\\
0 & 0 & 0 & -1
\end{array}\right)$}} & \multicolumn{1}{c}{\multirow{4}{*}{$\left(\begin{array}{cccc}
0 & \frac{1}{2} & \frac{1}{2} & 0\\
\frac{1}{2} & 0 & 0 & \frac{1}{2}\\
\frac{1}{2} & 0 & 0 & \frac{1}{2}\\
0 & \frac{1}{2} & \frac{1}{2} & 0
\end{array}\right)$}} & \multicolumn{1}{c}{\multirow{4}{*}{$\left(\begin{array}{cccc}
0 & -\frac{1}{2} & -\frac{1}{2} & 0\\
-\frac{1}{2} & 0 & 0 & -\frac{1}{2}\\
-\frac{1}{2} & 0 & 0 & -\frac{1}{2}\\
0 & -\frac{1}{2} & -\frac{1}{2} & 0
\end{array}\right)$}}\tabularnewline

 & \multicolumn{1}{c}{$\phi_{2}=\left[[QQ]_{\bar{3}_{c}}^{1}\left[[\bar{Q}\bar{Q}]_{3_{c}}^{0},\text{\ensuremath{\rho}}\right]_{\bar{3}_{c}}^{1}\right]^{0}$}
&  &  &  & \tabularnewline

 & \multicolumn{1}{c}{$\phi_{3}=\left[\left[[QQ]_{6_{c}}^{1},\rho\right]_{6_{c}}^{0}[\bar{Q}\bar{Q}]_{\bar{6}_{c}}^{0}\right]^{0}$}
&  &  &  & \tabularnewline

&
$\phi_{4}=\left[\left[QQ\right]_{6_{c}}^{0}\left[[\bar{Q}\bar{Q}]_{\bar{6}_{c}}^{1},\rho\right]_{\bar{6}_{c}}^{0}\right]^{0}$&
&  &  & \tabularnewline \hline
 \multirow{4}{*}{$1^{-}$} & \multicolumn{1}{c}{$\phi_{1}=\left[\left[[QQ]_{\bar{3}_{c}}^{0},\text{\ensuremath{\rho}}\right]_{\bar{3}_{c}}^{1}[\bar{Q}\bar{Q}]_{3_{c}}^{1}\right]^{1}$}  & \multicolumn{1}{c}{\multirow{4}{*}{$\left(\begin{array}{cccc}
0 & 0 & 0 & 0\\
0 & 0 & 0 & 0\\
0 & 0 & \text{\ensuremath{\frac{1}{2}}} & 0\\
0 & 0 & 0 & 0
\end{array}\right)$}} & \multicolumn{1}{c}{\multirow{4}{*}{$\left(\begin{array}{cccc}
0 & 0 & 0 & 0\\
0 & 0 & 0 & 0\\
0 & 0 & 0 & 0\\
0 & 0 & 0 & \text{\ensuremath{\frac{1}{2}}}
\end{array}\right)$}} & \multicolumn{1}{c}{\multirow{4}{*}{$\left(\begin{array}{cccc}
0 & -\frac{1}{4} & -\frac{1}{4} & 0\\
-\frac{1}{4} & 0 & 0 & -\frac{1}{4}\\
-\frac{1}{4} & 0 & 0 & -\frac{1}{4}\\
0 & -\frac{1}{4} & -\frac{1}{4} & 0
\end{array}\right)$}} & \multicolumn{1}{c}{\multirow{4}{*}{$\left(\begin{array}{cccc}
0 & \frac{1}{4} & \frac{1}{4} & 0\\
\frac{1}{4} & 0 & 0 & \frac{1}{4}\\
\frac{1}{4} & 0 & 0 & \frac{1}{4}\\
0 & \frac{1}{4} & \frac{1}{4} & 0
\end{array}\right)$}}\tabularnewline

  & \multicolumn{1}{c}{$\phi_{2}=\left[[QQ]_{\bar{3}_{c}}^{1}\left[[\bar{Q}\bar{Q}]_{3_{c}}^{0},\text{\ensuremath{\rho}}\right]_{\bar{3}_{c}}^{1}\right]^{1}$} &  &  &  & \tabularnewline

 &  \multicolumn{1}{c}{$\phi_{3}=\left[\left[[QQ]_{6_{c}}^{1},\rho\right]_{6_{c}}^{1}[\bar{Q}\bar{Q}]_{\bar{6}_{c}}^{0}\right]^{1}$} &  &  &  & \tabularnewline

&
$\phi_{4}=\left[\left[QQ\right]_{6_{c}}^{0}\left[[\bar{Q}\bar{Q}]_{\bar{6}_{c}}^{1},\rho\right]_{\bar{6}_{c}}^{1}\right]^{1}$
&  &  &  & \tabularnewline \hline \multirow{4}{*}{$2^{-}$} &
\multicolumn{1}{c}{
$\phi_{1}=\left[\left[[QQ]_{\bar{3}_{c}}^{0},\text{\ensuremath{\rho}}\right]_{\bar{3}_{c}}^{1}[\bar{Q}\bar{Q}]_{3_{c}}^{1}\right]^{2}
$}   & \multicolumn{1}{c}{\multirow{3}{*}{$\left(\begin{array}{cccc}
0 & 0 & 0 & 0\\
0 & 0 & 0 & 0\\
0 & 0 & \text{\ensuremath{-\frac{1}{10}}} & 0\\
0 & 0 & 0 & 0
\end{array}\right)$}} & \multicolumn{1}{c}{\multirow{3}{*}{$\left(\begin{array}{cccc}
0 & 0 & 0 & 0\\
0 & 0 & 0 & 0\\
0 & 0 & 0 & 0\\
0 & 0 & 0 & \text{-\ensuremath{\frac{1}{10}}}
\end{array}\right)$}} & \multicolumn{1}{c}{\multirow{3}{*}{$\left(\begin{array}{cccc}
0 & \frac{1}{20} & \frac{1}{20} & 0\\
\frac{1}{20} & 0 & 0 & \frac{1}{20}\\
\frac{1}{20} & 0 & 0 & \frac{1}{20}\\
0 & \frac{1}{20} & \frac{1}{20} & 0
\end{array}\right)$}} & \multicolumn{1}{c}{\multirow{3}{*}{$\left(\begin{array}{cccc}
0 & -\frac{1}{20} & -\frac{1}{20} & 0\\
-\frac{1}{20} & 0 & 0 & -\frac{1}{20}\\
-\frac{1}{20} & 0 & 0 & -\frac{1}{20}\\
0 & -\frac{1}{20} & -\frac{1}{20} & 0
\end{array}\right)$}}\tabularnewline

 & \multicolumn{1}{c}{$\phi_{2}=\left[[QQ]_{\bar{3}_{c}}^{1}\left[[\bar{Q}\bar{Q}]_{3_{c}}^{0},\text{\ensuremath{\rho}}\right]_{\bar{3}_{c}}^{1}\right]^{2}$} &  &  &  & \tabularnewline

 & \multicolumn{1}{c}{$\phi_{3}=\left[\left[[QQ]_{6_{c}}^{1},\rho\right]_{6_{c}}^{2}[\bar{Q}\bar{Q}]_{\bar{6}_{c}}^{0}\right]^{2}$}  &  &  &  & \tabularnewline

&
\multicolumn{1}{c}{$\phi_{4}=\left[\left[QQ\right]_{6_{c}}^{0}\left[[\bar
{Q}\bar
{Q}]_{\bar{6}_{c}}^{1},\rho\right]_{\bar{6}_{c}}^{2}\right]^{2}$} &
&  &  & \tabularnewline \bottomrule[1pt]
\end{tabular}}
\end{table*}

\begin{table*}
 \renewcommand\arraystretch{1.5}
  \caption{The spin-orbital and tensor factors for the mixing of the $\lambda$-mode and $\rho$-mode P-wave states. For conciseness, we list the unit for each subregion in the table and the factor is the product of the unit and $0, \pm 1$.  }\label{lolambdarho}
 \centering
 \setlength{\tabcolsep}{2.0mm}
   \resizebox{\textwidth}{11cm}{
\begin{tabular}{c|c|cccccc|cccccc}
    \toprule[1pt]
    $J^{PC}$ &  & \multicolumn{6}{c}{$\mathbf{L}_{ij}\cdot\mathbf{S}_{ij}$} & \multicolumn{6}{c}{$\mathcal{S}_{ij}$}\tabularnewline
\hline
    &  & $Q_{1}Q_{2}$ & $\bar{Q}_{3}\bar{Q}_{4}$ & $Q_{1}\bar{Q}_{3}$ & $Q_{2}\bar{Q}_{4}$ & $Q_{1}\bar{Q}_{4}$ & $Q_{2}\bar{Q}_{3}$ & $Q_{1}Q_{2}$ & $\bar{Q}_{3}\bar{Q}_{4}$ & $Q_{1}\bar{Q}_{3}$ & $Q_{2}\bar{Q}_{4}$ & $Q_{1}\bar{Q}_{4}$ & $Q_{2}\bar{Q}_{3}$\tabularnewline
\hline

    &  & \multicolumn{6}{c|}{$\left[\left[[QQ]_{\bar{3}_{c}}^{1}[\bar{Q}\bar{Q}]_{3_{c}}^{1}\right]_{1_{c}}^{1},\lambda\right]_{1_{c}}^{0}$(unit
        of $\frac{1}{\sqrt{2}}$)} & \multicolumn{6}{c}{$\left[\left[[QQ]_{\bar{3}_{c}}^{1}[\bar{Q}\bar{Q}]_{3_{c}}^{1}\right]_{1_{c}}^{1},\lambda\right]_{1_{c}}^{0}$(unit
        of $\frac{1}{2\sqrt{2}}$)}\tabularnewline
    \multirow{3}{*}{$0^{-+}$} & $\left[\left[[QQ]_{\bar{3}_{c}}^{0},\text{\ensuremath{\rho}}\right]_{\bar{3}_{c}}^{1}[\bar{Q}\bar{Q}]_{3_{c}}^{1}\right]^{0}$ & $0$ & $0$ & $1$ & $-1$ & $1$ & $-1$ & $0$ & $0$ & $1$ & $-1$ & $1$ & $-1$\tabularnewline
    & $\left[[QQ]_{\bar{3}_{c}}^{1}\left[[\bar{Q}\bar{Q}]_{3_{c}}^{0},\text{\ensuremath{\rho}}\right]_{\bar{3}_{c}}^{1}\right]^{0}$ & $0$ & $0$ & $-1$ & $1$ & $1$ & $-1$ & $0$ & $0$ & $-1$ & $1$ & $1$ & $-1$\tabularnewline
    & $\left[\left[[QQ]_{6_{c}}^{1},\rho\right]_{6_{c}}^{0}[\bar{Q}\bar{Q}]_{\bar{6}_{c}}^{0}\right]^{0}$ & $0$ & $0$ & $-1$ & $1$ & $1$ & $-1$ & $0$ & $0$ & $-1$ & $1$ & $1$ & $-1$\tabularnewline
    & $\left[\left[QQ\right]_{6_{c}}^{0}\left[[\bar{Q}\bar{Q}]_{\bar{6}_{c}}^{1},\rho\right]_{\bar{6}_{c}}^{0}\right]^{0}$ & $0$ & $0$ & $1$ & $-1$ & $1$ & $-1$ & $0$ & $0$ & $1$ & $-1$ & $1$ & $-1$\tabularnewline
\hline
    \multirow{5}{*}{$1^{-+}$} &  & \multicolumn{6}{c|}{$\left[\left[[QQ]_{\bar{3}_{c}}^{1}[\bar{Q}\bar{Q}]_{3_{c}}^{1}\right]_{1_{c}}^{1},\lambda\right]_{1_{c}}^{1}$(unit
        of $\frac{1}{2\sqrt{2}}$)} & \multicolumn{6}{c}{$\left[\left[[QQ]_{\bar{3}_{c}}^{1}[\bar{Q}\bar{Q}]_{3_{c}}^{1}\right]_{1_{c}}^{1},\lambda\right]_{1_{c}}^{1}$(unit
        of $\frac{1}{4\sqrt{2}}$)}\tabularnewline
    & $\left[\left[[QQ]_{\bar{3}_{c}}^{0},\text{\ensuremath{\rho}}\right]_{\bar{3}_{c}}^{1}[\bar{Q}\bar{Q}]_{3_{c}}^{1}\right]^{1}$ & $0$ & $0$ & $1$ & $-1$ & $1$ & $-1$ & $0$ & $0$ & $-1$ & $1$ & $-1$ & $1$\tabularnewline
    & $\left[[QQ]_{\bar{3}_{c}}^{1}\left[[\bar{Q}\bar{Q}]_{3_{c}}^{0},\text{\ensuremath{\rho}}\right]_{\bar{3}_{c}}^{1}\right]^{1}$ & $0$ & $0$ & $-1$ & $1$ & $1$ & $-1$ & $0$ & $0$ & $1$ & $-1$ & $-1$ & $1$\tabularnewline
    & $\left[\left[[QQ]_{6_{c}}^{1},\rho\right]_{6_{c}}^{1}[\bar{Q}\bar{Q}]_{\bar{6}_{c}}^{0}\right]^{1}$ & $0$ & $0$ & $-1$ & $1$ & $1$ & $-1$ & $0$ & $0$ & $1$ & $-1$ & $-1$ & $1$\tabularnewline
    & $\left[\left[QQ\right]_{6_{c}}^{0}\left[[\bar{Q}\bar{Q}]_{\bar{6}_{c}}^{1},\rho\right]_{\bar{6}_{c}}^{1}\right]^{1}$ & $0$ & $0$ & $1$ & $-1$ & $1$ & $-1$ & $0$ & $0$ & $-1$ & $1$ & $-1$ & $1$\tabularnewline
\hline
    \multirow{15}{*}{$1^{--}$} &  & \multicolumn{6}{c|}{$\left[\left[[QQ]_{\bar{3}_{c}}^{1}[\bar{Q}\bar{Q}]_{3_{c}}^{1}\right]_{1_{c}}^{0},\lambda\right]_{1_{c}}^{1}$(unit
        of $\frac{1}{\sqrt{6}}$)} & \multicolumn{6}{c}{$\left[\left[[QQ]_{\bar{3}_{c}}^{1}[\bar{Q}\bar{Q}]_{3_{c}}^{1}\right]_{1_{c}}^{0},\lambda\right]_{1_{c}}^{1}$}\tabularnewline
    & $\left[\left[[QQ]_{\bar{3}_{c}}^{0},\text{\ensuremath{\rho}}\right]_{\bar{3}_{c}}^{1}[\bar{Q}\bar{Q}]_{3_{c}}^{1}\right]^{1}$ & $0$ & $0$ & $1$ & $-1$ & $1$ & $-1$ & $0$ & $0$ & $0$ & $0$ & $0$ & $0$\tabularnewline
    & $\left[[QQ]_{\bar{3}_{c}}^{1}\left[[\bar{Q}\bar{Q}]_{3_{c}}^{0},\text{\ensuremath{\rho}}\right]_{\bar{3}_{c}}^{1}\right]^{1}$ & $0$ & $0$ & $1$ & $-1$ & $-1$ & $1$ & $0$ & $0$ & $0$ & $0$ & $0$ & $0$\tabularnewline
    & $\left[\left[[QQ]_{6_{c}}^{1},\rho\right]_{6_{c}}^{1}[\bar{Q}\bar{Q}]_{\bar{6}_{c}}^{0}\right]^{1}$ & $0$ & $0$ & $1$ & $-1$ & $-1$ & $1$ & $0$ & $0$ & $0$ & $0$ & $0$ & $0$\tabularnewline
    & $\left[\left[QQ\right]_{6_{c}}^{0}\left[[\bar{Q}\bar{Q}]_{\bar{6}_{c}}^{1},\rho\right]_{\bar{6}_{c}}^{1}\right]^{1}$ & $0$ & $0$ & $1$ & $-1$ & $1$ & $-1$ & $0$ & $0$ & $0$ & $0$ & $0$ & $0$\tabularnewline
    &  & \multicolumn{6}{c|}{$\left[\left[[QQ]_{6_{c}}^{0}[\bar{Q}\bar{Q}]_{\bar{6}_{c}}^{0}\right]_{1_{c}}^{0},\lambda\right]_{1_{c}}^{1}$(unit
        of $\frac{1}{\sqrt{2}}$)} & \multicolumn{6}{c}{$\left[\left[[QQ]_{6_{c}}^{0}[\bar{Q}\bar{Q}]_{\bar{6}_{c}}^{0}\right]_{1_{c}}^{0},\lambda\right]_{1_{c}}^{1}$}\tabularnewline
    & $\left[\left[[QQ]_{\bar{3}_{c}}^{0},\text{\ensuremath{\rho}}\right]_{\bar{3}_{c}}^{1}[\bar{Q}\bar{Q}]_{3_{c}}^{1}\right]^{1}$ & $0$ & $0$ & $-1$ & $1$ & $1$ & $-1$ & $0$ & $0$ & $0$ & $0$ & $0$ & $0$\tabularnewline
    & $\left[[QQ]_{\bar{3}_{c}}^{1}\left[[\bar{Q}\bar{Q}]_{3_{c}}^{0},\text{\ensuremath{\rho}}\right]_{\bar{3}_{c}}^{1}\right]^{1}$ & $0$ & $0$ & $-1$ & $1$ & $-1$ & $1$ & $0$ & $0$ & $0$ & $0$ & $0$ & $0$\tabularnewline
    & $\left[\left[[QQ]_{6_{c}}^{1},\rho\right]_{6_{c}}^{1}[\bar{Q}\bar{Q}]_{\bar{6}_{c}}^{0}\right]^{1}$ & $0$ & $0$ & $-1$ & $1$ & $-1$ & $1$ & $0$ & $0$ & $0$ & $0$ & $0$ & $0$\tabularnewline
    & $\left[\left[QQ\right]_{6_{c}}^{0}\left[[\bar{Q}\bar{Q}]_{\bar{6}_{c}}^{1},\rho\right]_{\bar{6}_{c}}^{1}\right]^{1}$ & $0$ & $0$ & $-1$ & $1$ & $1$ & $-1$ & $0$ & $0$ & $0$ & $0$ & $0$ & $0$\tabularnewline
    \cmidrule{2-14} \cmidrule{3-14} \cmidrule{4-14} \cmidrule{5-14} \cmidrule{6-14} \cmidrule{7-14} \cmidrule{8-14} \cmidrule{9-14} \cmidrule{10-14} \cmidrule{11-14} \cmidrule{12-14} \cmidrule{13-14} \cmidrule{14-14}
    &  & \multicolumn{6}{c|}{$\left[\left[[QQ]_{\bar{3}_{c}}^{1}[\bar{Q}\bar{Q}]_{3_{c}}^{1}\right]_{1_{c}}^{2},\lambda\right]_{1_{c}}^{1}$(unit
        of $\frac{\sqrt{30}}{12}$)} & \multicolumn{6}{c}{$\left[\left[[QQ]_{\bar{3}_{c}}^{1}[\bar{Q}\bar{Q}]_{3_{c}}^{1}\right]_{1_{c}}^{2},\lambda\right]_{1_{c}}^{1}$(unit
        of $\frac{3\sqrt{30}}{40}$)}\tabularnewline
    & $\left[\left[[QQ]_{\bar{3}_{c}}^{0},\text{\ensuremath{\rho}}\right]_{\bar{3}_{c}}^{1}[\bar{Q}\bar{Q}]_{3_{c}}^{1}\right]^{1}$ & $0$ & $0$ & $-1$ & $1$ & $-1$ & $1$ & $0$ & $0$ & $-1$ & $1$ & $-1$ & $1$\tabularnewline
    & $\left[[QQ]_{\bar{3}_{c}}^{1}\left[[\bar{Q}\bar{Q}]_{3_{c}}^{0},\text{\ensuremath{\rho}}\right]_{\bar{3}_{c}}^{1}\right]^{1}$ & $0$ & $0$ & $-1$ & $1$ & $1$ & $-1$ & $0$ & $0$ & $-1$ & $1$ & $1$ & $-1$\tabularnewline
    & $\left[\left[[QQ]_{6_{c}}^{1},\rho\right]_{6_{c}}^{1}[\bar{Q}\bar{Q}]_{\bar{6}_{c}}^{0}\right]^{1}$ & $0$ & $0$ & $-1$ & $1$ & $1$ & $-1$ & $0$ & $0$ & $-1$ & $1$ & $1$ & $-1$\tabularnewline
    & $\left[\left[QQ\right]_{6_{c}}^{0}\left[[\bar{Q}\bar{Q}]_{\bar{6}_{c}}^{1},\rho\right]_{\bar{6}_{c}}^{1}\right]^{1}$ & $0$ & $0$ & $-1$ & $1$ & $-1$ & $1$ & $0$ & $0$ & $-1$ & $1$ & $-1$ & $1$\tabularnewline
\hline
    \multirow{5}{*}{$2^{-+}$} &  & \multicolumn{6}{c|}{$\left[\left[[QQ]_{\bar{3}_{c}}^{1}[\bar{Q}\bar{Q}]_{3_{c}}^{1}\right]_{1_{c}}^{1},\lambda\right]_{1_{c}}^{0}$
        (unit of $\frac{1}{2\sqrt{2}}$)} & \multicolumn{6}{c}{$\left[\left[[QQ]_{\bar{3}_{c}}^{1}[\bar{Q}\bar{Q}]_{3_{c}}^{1}\right]_{1_{c}}^{1},\lambda\right]_{1_{c}}^{0}$
        (unit of $\frac{1}{20\sqrt{2}}$)}\tabularnewline
    & $\left[\left[[QQ]_{\bar{3}_{c}}^{0},\text{\ensuremath{\rho}}\right]_{\bar{3}_{c}}^{1}[\bar{Q}\bar{Q}]_{3_{c}}^{1}\right]^{2}$ & $0$ & $0$ & $-1$ & $1$ & $-1$ & $1$ & $0$ & $0$ & $1$ & $-1$ & $1$ & $-1$\tabularnewline
    & $\left[[QQ]_{\bar{3}_{c}}^{1}\left[[\bar{Q}\bar{Q}]_{3_{c}}^{0},\text{\ensuremath{\rho}}\right]_{\bar{3}_{c}}^{1}\right]^{2}$ & $0$ & $0$ & $1$ & $-1$ & $-1$ & $1$ & $0$ & $0$ & $-1$ & $1$ & $1$ & $-1$\tabularnewline
    & $\left[\left[[QQ]_{6_{c}}^{1},\rho\right]_{6_{c}}^{2}[\bar{Q}\bar{Q}]_{\bar{6}_{c}}^{0}\right]^{2}$ & $0$ & $0$ & $1$ & $-1$ & $-1$ & $1$ & $0$ & $0$ & $-1$ & $1$ & $1$ & $-1$\tabularnewline
    & $\left[\left[QQ\right]_{6_{c}}^{0}\left[[\bar{Q}\bar{Q}]_{\bar{6}_{c}}^{1},\rho\right]_{\bar{6}_{c}}^{2}\right]^{2}$ & $0$ & $0$ & $-1$ & $1$ & $-1$ & $1$ & $0$ & $0$ & $1$ & $-1$ & $1$ & $-1$\tabularnewline
\hline
    \multirow{5}{*}{$2^{--}$} &  & \multicolumn{6}{c|}{$\left[\left[[QQ]_{\bar{3}_{c}}^{1}[\bar{Q}\bar{Q}]_{3_{c}}^{1}\right]_{1_{c}}^{2},\lambda\right]_{1_{c}}^{2}$
        (unit of $\frac{\sqrt{6}}{4}$)} & \multicolumn{6}{c}{$\left[\left[[QQ]_{\bar{3}_{c}}^{1}[\bar{Q}\bar{Q}]_{3_{c}}^{1}\right]_{1_{c}}^{2},\lambda\right]_{1_{c}}^{2}$
        (unit of $\frac{\sqrt{6}}{4}$)}\tabularnewline
    & $\left[\left[[QQ]_{\bar{3}_{c}}^{0},\text{\ensuremath{\rho}}\right]_{\bar{3}_{c}}^{1}[\bar{Q}\bar{Q}]_{3_{c}}^{1}\right]^{2}$ & $0$ & $0$ & $1$ & $-1$ & $1$ & $-1$ & $0$ & $0$ & $-1$ & $1$ & $-1$ & $1$\tabularnewline
    & $\left[[QQ]_{\bar{3}_{c}}^{1}\left[[\bar{Q}\bar{Q}]_{3_{c}}^{0},\text{\ensuremath{\rho}}\right]_{\bar{3}_{c}}^{1}\right]^{2}$ & $0$ & $0$ & $1$ & $-1$ & $-1$ & $1$ & $0$ & $0$ & $-1$ & $1$ & $1$ & $-1$\tabularnewline
    & $\left[\left[[QQ]_{6_{c}}^{1},\rho\right]_{6_{c}}^{2}[\bar{Q}\bar{Q}]_{\bar{6}_{c}}^{0}\right]^{2}$ & $0$ & $0$ & $1$ & $-1$ & $-1$ & $1$ & $0$ & $0$ & $-1$ & $1$ & $1$ & $-1$\tabularnewline
    & $\left[\left[QQ\right]_{6_{c}}^{0}\left[[\bar{Q}\bar{Q}]_{\bar{6}_{c}}^{1},\rho\right]_{\bar{6}_{c}}^{2}\right]^{2}$ & $0$ & $0$ & $1$ & $-1$ & $1$ & $-1$ & $0$ & $0$ & $-1$ & $1$ & $-1$ & $1$\tabularnewline
    \bottomrule[1pt]
\end{tabular}}
\end{table*}

\begin{acknowledgements}
We are grateful for the helpful discussions with Xin-Zhen Weng. This
project was supported by the National Natural Science Foundation of
China (11975033 and 12070131001). M.~Oka is supported in part by the
JSPS KAKENHI (Nos.~19H05159, 20K03959, and 21H00132). G.J.~Wang was
supported by JSPS KAKENHI (No. 20F20026).~L.~Meng was funded by the
Deutsche Forschungsgemeinschaft (DFG, German Research Foundation,
Project ID 196253076-TRR 110).
\end{acknowledgements}

\bibliography{ref}

\end{document}